\documentclass[fleqn,usenatbib]{mnras}

\usepackage{mathptmx}

\usepackage[T1]{fontenc}

\DeclareRobustCommand{\VAN}[3]{#2}
\let\VANthebibliography\thebibliography
\def\thebibliography{\DeclareRobustCommand{\VAN}[3]{##3}\VANthebibliography}

\usepackage{graphicx}	
\usepackage{amsmath}	
\usepackage{amssymb}	

\usepackage{xcolor}    

\newcommand{\sigmasfr}{$\Sigma_{\rm SFR}$}
\newcommand{\msun}{M$_\odot$}
\newcommand{\sigmasfrunits}{\msun~yr$^{-1}$~kpc$^{-2}$}
\newcommand{\vout}{$v_{\rm out}$}

\title[Resolved Outflows in a Starbursting Galaxy]{The DUVET Survey: Resolved Maps of Star Formation Driven Outflows in a Compact, Starbursting Disk Galaxy}

\author[B. Reichardt Chu et al.]{Bronwyn Reichardt Chu,$^{1,2}$\thanks{E-mail: breichardtchu@swin.edu.au}
Deanne B. Fisher,$^{1,2}$
Nikole M. Nielsen,$^{1,2}$ John Chisholm,$^{3,4}$   
\newauthor{Marianne Girard,$^{1,2}$ Glenn G. Kacprzak,$^{1,2}$ Alberto Bolatto,$^{5}$ Rodrigo Herrera-Camus,$^{6}$ }
\newauthor{Karin Sandstrom,$^{7}$ Miao Li,$^{8,9}$ Ryan Rickards Vaught$^{7}$
and Daniel K. McPherson$^{1,2}$}
\\
$^{1}$Centre for Astrophysics and Supercomputing, Swinburne University of Technology, Hawthorn, VIC 3122, Australia\\
$^{2}$ARC Centre of Excellence for All Sky Astrophysics in 3 Dimensions (ASTRO 3D)\\
$^{3}$Department of Astronomy, University of Texas, Austin, TX, USA\\
$^{4}$Hubble Fellow\\
$^{5}$University of Maryland, College Park, MD, USA\\
$^{6}$Departamento de Astronom\'ia, Universidad de Concepci\'on, Barrio Universitario, Concepci\'on, Chile\\
$^{7}$Center for Astrophysics and Space Sciences, Department of Physics, University of California, San Diego, CA, USA\\
$^{8}$Department of Physics, Zhejiang University, 866 Yuhangtang Road, Hangzhou, 310058, China\\
$^{9}$Center for Computational Astrophysics, Flatiron Institute, 162 Fifth Avenue, New York, NY 10010, USA
}

\date{Accepted 2022 February 5. Received 2022 February 3; in original form 2021 June 25}

\pubyear{2022}

\begin{document}
\label{firstpage}
\pagerange{\pageref{firstpage}--\pageref{lastpage}}
\maketitle

\begin{abstract}
We study star formation driven outflows in a $z\sim0.02$ starbursting disk galaxy, IRAS08339+6517, using spatially resolved measurements from the Keck Cosmic Web Imager (KCWI).  We develop a new method incorporating a multi-step process to determine whether an outflow should be fit in each spaxel, and then subsequently decompose the emission line into multiple components. We detect outflows ranging in velocity, \vout, from $100-600$~km~s$^{-1}$ across a range of star formation rate surface densities, \sigmasfr, from $\sim$0.01-10~\sigmasfrunits\ in resolution elements of a few hundred parsec. Outflows are detected in $\sim$100\% of all spaxels within the half-light radius,and $\sim70$\% within $r_{90}$, suggestive of a high covering fraction for this starbursting disk galaxy. Around $2/3$ of the total outflowing mass originates from the star forming ring, which corresponds to $<10\%$ of the total area of the galaxy.  
We find that the relationship between \vout\ and the \sigmasfr, as well as between the mass loading factor, $\eta$, and the \sigmasfr, are consistent with trends expected from energy-driven feedback models. We study the resolution effects on this relationship and find stronger correlations above a re-binned size-scale of $\sim500$~pc. Conversely, we do not find statistically significant consistency with the prediction from momentum-driven winds. 
\end{abstract}




\section{Introduction} \label{sec:intro}

Galactic outflows have been observed ubiquitously in star-forming galaxies across cosmic time \citep{heckman2000absorptionline, chen2010absorption, Rubin2014evidence}.  
Outflows driven by star formation are thought to play an integral role in the evolution of galaxies \citep[e.g.][]{tumlinson2017CGMreview,veilleux2005galacticwinds, veilleux2020cooloutflows}.  Outflows are an observational signature of the feedback process which is thought to regulate star formation within galaxies \citep[e.g.][]{ostriker2010regulation}.  When outflows are not included, simulations fail to reproduce basic properties of galaxies such as the galaxy mass function, typical galaxy sizes, and the Kennicutt-Schmidt Law \citep[e.g][]{springel2003cosmologicalsim, oppenheimer2006cosmologicalsims, hopkins2012stellarfeedback, hopkins2014FIRE}. Characterising the properties of outflows in star-forming galaxies is therefore a critical goal of extragalactic astronomy.\footnote{We note that in this paper we refer to ``outflows" as the movement of gas outwards from a source within the disk of the galaxy.  We do not require that gas be above the escape velocity to be called an outflow. }

A popular theory of star formation is that in star-forming disks the internal gravitational potential is balanced by pressure from turbulence driven by feedback from star formation, generating a self-regulating process \citep{ostriker2010regulation}.  In such theories, the star formation rate surface density, $\Sigma_{\mathrm{SFR}}$, is linearly proportional to the midplane pressure, $P$, of the galaxy disk \citep{Shetty2012maximally,kim2013threeDhydrosim}.  However, \cite{fisher2019testing} found that while a strong positive correlation existed between \sigmasfr\ and P across almost 6 orders of magnitude, in galaxies with \sigmasfr~$> 0.1$~\sigmasfrunits\ the values of $P/\Sigma_{\mathrm{SFR}}$ were more than an order of magnitude higher than predicted \citep[see also][]{girard2021systematic}.  Moreover, \citet{krumholz2018unifiedmodel} argued that star formation feedback alone is not sufficient to generate the high velocity dispersions in star bursting disk galaxies. \cite{girard2021systematic} showed that this remains true when measuring velocity dispersion with molecular gas \citep[see also][]{wilson2019kennicuttschmidtlaw}. In short, the feedback models which do well to describe the properties of Milky Way like environments do not generate enough turbulent support to match the properties in highly star-forming disk galaxies. 

The feedback-regulated models described above make the assumption that turbulence is primarily due to mechanical energy from supernova (energy-driven feedback), however \cite{murray2011radiationpressure} put forward the idea that radiative pressure from young stars could drive turbulence as well (momentum-driven feedback). Simulations that incorporate the momentum-driven launching mechanism to drive outflows do produce significantly higher gas velocity dispersions \citep{hung2019whatdrives}. We note however, that recent work which used molecular gas as a tracer of velocity dispersions found that while the velocity dispersion of the energy-driven wind model remained too low, that of the momentum-driven winds may be, in fact, too high \citep{ubler2019evolutionandorigin, girard2021systematic}.    

Each model has different predictions for the relationship between the outflow velocity \vout\ and \sigmasfr, and the mass loading factor $\eta$ and \sigmasfr.  
Energy-driven outflows depend on the star formation activity occurring within the galaxy as $v_{out}\propto$~\sigmasfr$^{0.1}$ \citep{chen2010absorption,li2017supernovaedriven,kim2020firstresultssmaug}, or $v_{out}\propto\mathrm{SFR}^{0.2}$ \citep{ferrara2006winds}. Recent simulations based on observations of ionised gas outflows and using energy-driven feedback have found a relationship for the mass loading factor of $\eta\propto$~\sigmasfr$^{-0.44}$ \citep{li2017supernovaedriven, kim2020firstresultssmaug, li2020simpleyetpowerful}. On the other hand, momentum-driven outflows have a steeper dependency on star formation activity, with $v_{out}\propto$~\sigmasfr$^2$ \citep{murray2011radiationpressure}, or $v_{out}\propto\mathrm{SFR}$ \citep{hopkins2012stellarfeedback}. We can therefore use observations of outflows to directly test the models of the feedback launching mechanism. 


A challenge to studying outflows is that their contribution is much fainter than the signal from the rest of the galaxy. The typical emission line outflow is oftentimes $\sim75\%$ of the bright nebular emission from areas with star-formation within the galaxy \citep{newman2012sins, arribas2014ionisedgasoutflows, davies2019kiloparsec}. For areas with \sigmasfr~$<0.2$~\sigmasfrunits, the typical emission from the outflow can be $<15\%$ of the flux of the bright nebular emission \citep{davies2019kiloparsec}. To combat this, the majority of observational studies of star formation driven outflows have focused on the integrated light from entire galaxy measurements. It is however difficult to test the correlations which will constrain the main driving mechanisms of star formation feedback with galaxy-wide observations as there is a well-known underlying dependence on the stellar mass of the galaxy such that more massive galaxies drive outflows with higher velocities \citep{newman2012sins, chisholm2015scaling, nelson2019firstresultsTNG50}. 

Galaxy stacks have been used to increase the signal-to-noise such that the signature of star formation driven outflows can be distinguished from the galaxy emission or absorption \citep[e.g.][]{weiner2009ubiquitous, chen2010absorption, davies2019kiloparsec}.  However stacking spectra removes important details \citep{jones2018dust} such as the velocity dispersion, and large numbers of galaxies must be included in the sample.

Various studies have reported trends of outflow velocity with \sigmasfr\ \citep{chisholm2015scaling, newman2012shocked, davies2019kiloparsec}, stellar mass \citep{chisholm2015scaling, newman2012sins}, and extinction \citep{chen2010absorption}.   The slope of the \vout\ to star formation rate (SFR) relationship has been observed by some studies to be in the range expected for an energy-driven outflow, with slopes of 0.1-0.15 \citep[e.g.][]{chen2010absorption, arribas2014ionisedgasoutflows, chisholm2015scaling}.  Further studies found steeper slopes of up to 0.35 \citep[e.g.][]{martin2005mapping, rupke2005outflowsdiscussion, heckman2016implications, davies2019kiloparsec} which are still not in the expected range for momentum-driven outflows. Overall there is a lack of consensus around the nature of the relationship between outflow velocity and \sigmasfr\ which may be related to systematics of the measurement methods.

In this paper we study a local galaxy with high \sigmasfr\ in order to spatially resolve the outflow measurements. We develop a multi-step method for automatically identifying outflows in an IFU data set.  We remove the dependence of the outflow velocity on total galaxy stellar mass while mapping any changes in the relationship between outflow kinematics, mass loading factor and local galaxy properties across the face of our target starbursting disk galaxy.  This allows us to isolate the SFR as the only dependent variable to test what drives the outflow properties.

The paper is organised as follows. We describe our target galaxy (IRAS08339+6517) and the design of the DUVET survey in Section \ref{sec:pilot_target}.  Our observations and data reduction of our pilot target are described in Section \ref{sec:observations} and are used together with our new method to measure spatially resolved outflows in Section \ref{sec:koffee}.  In Section \ref{sec:results} we explore the relationship between the outflow velocity, the broad-to-narrow flux ratio and the mass loading factor with star formation rate surface density, and map the distribution of outflows across the face of IRAS08339+6517.  Our discussion of our results and conclusions are presented in Section \ref{sec:summary}.

We assume a flat $\Lambda$CDM cosmology with $H_0~=~69.3~\mathrm{km}~\mathrm{Mpc}^{-1}~\mathrm{s}^{-1}$ and $\Omega_0=0.3$ \citep{hinshaw2013wmap9}.

\section{DUVET Pilot Target: IRAS08339+6517} \label{sec:pilot_target}

\subsection{DUVET Project}
These observations represent the pilot program for the DUVET survey (Fisher et al. {\em in prep}). DUVET (Deep near-UV observations of Entrained gas in Turbulent galaxies) is a survey of 27 starbursting galaxies at $z\approx0.02-0.04$ aimed at studying star formation feedback in high \sigmasfr\ environments.  One of the goals of the survey is to be able to measure kinematics across the face of the galaxies, and use these kinematics to study the outflowing gas (as traced by broad lines discussed in this paper) and its driving mechanisms. The main criteria for the sample is that the galaxies must have a SFR that is at minimum 5$\times$ the main-sequence value for the corresponding stellar mass. Secondly, they must have disk morphology and kinematics. Minor-merger interactions are present in many of the DUVET targets, however we ensure that this does not dominate the morpho-kinematic state of the system. We define ``disks" as galaxies with (1) exponential surface brightness profiles in $i$-band SDSS images; and (2) velocity fields that are dominated by rotation and have a well defined kinematic centre. This definition is similar to large surveys of starbursting galaxies at $z\approx1-2$ \citep{forsterschreiber2011sinssurvey, wisnioski2015KMOS3Dsurvey}. We also favour targets that are face-on. The mass range of the sample is 10$^9$-10$^{11}$~M$_{\odot}$, and has a range of metallicities from 0.1-1.5~Z$_{\odot}$. Future work will also investigate the stellar populations and gas properties via faint emission line features, such as [OIII]~$\lambda$4363.

\subsection{Pilot Target: IRAS 08339+6517}
IRAS 08339+6517 (hereafter IRAS08) is a well studied blue-compact galaxy at $z\approx~0.0191$. IRAS08 is face-on (inclination $\sim15^{\circ}-20^{\circ}$), bright in the UV, and has a young stellar population \citep{Leitherer2002globalFUV, lopezsanchez2006IRAS08paper, ostlin2009lymanalphamorphology, otifloranes2014physicalpropertiesIRAS08}. It is an ideal case for our study, as it has a well ordered rotation field and has a SFR roughly 10$\times$ greater than the galaxy main-sequence. It is therefore interesting for probing strong wind environments and easier to decompose the outflow component. 

IRAS08 has been treated in the literature as a galaxy with many properties similar to main-sequence galaxies at $z\approx 1-2$. Galaxies at redshifts $1-2$ are typically compact, starbursting, gas rich and clumpy \citep{madau2014cosmicSFH, genzel2011sins, tacconi2018PHIBSSunified}. IRAS08 has a half-light radius of 2.6\arcsec\ (1~kpc) in the $B$-band \citep{lopezsanchez2006IRAS08paper}, which is far more compact than typical in galaxies at redshift~$\sim~0$ \citep{mosleh2013robustness}.  The total stellar mass of IRAS08 is $M_*~\sim~10^{10}~\rm{M}_\odot$ \citep{lopezsanchez2006IRAS08paper}.  The SFR has been measured to be between 8-12~M$_\odot$~yr$^{-1}$ using a range of tracers \citep{gonzalezdelgado1998FUVspectra, lopezsanchez2006IRAS08paper}.  The galaxy averaged \sigmasfr\ of IRAS08 is $\sim$0.7~\sigmasfrunits, which is an order of magnitude higher than typical spiral galaxies of the local Universe \citep{saintonge2017xColdGass}. \citet{ostlin2009lymanalphamorphology} observed bright knots of star formation in the UV aligned along an internal bar \citep[see also][]{fisher2017DYNAMO-HSTsurvey}. These knots are similar to the so-called ``clumps" observed commonly in high-z galaxies \citep[e.g.][]{genzel2011sins}.  

Studies of IRAS08 have presented observations of tidal interactions with its companion dwarf galaxy 56~kpc away \citep{cannon2004extendedtidalstructure, lopezsanchez2006IRAS08paper}.  An HI tidal tail extends between the two galaxies, containing gas which has been stripped during the interaction \citep{cannon2004extendedtidalstructure, lopezsanchez2006IRAS08paper}. While these interactions may have initiated the current starburst event, they have not disturbed IRAS08's kinematics, such that it is still well-modelled as a rotating disk \citep{cannon2004extendedtidalstructure}.

IRAS08's properties make it an ideal target in which to study the mechanisms of star formation feedback in starbursting environments at resolved scales.

\section{Observations and Data Reduction}
\label{sec:observations}

IRAS08 was observed using the Keck Cosmic Web Imager \citep[KCWI,][]{Morrissey2018kcwi} on the Keck II telescope during the night of the 15th of February 2018 (UT).  Observations were taken using the large slicer (FOV: $33\arcsec~\times~20.4\arcsec$) with a position angle of $90^\circ$.  With this configuration, KCWI's spaxels are $0.3\arcsec\times1.35\arcsec$ ($0.1~\rm{kpc}\times0.5~\rm{kpc}$ at the distance of $\sim82~\mathrm{Mpc}$).  The blue BM grating was used with two central wavelength settings at $\lambda$=4050\AA\ and $\lambda$=4800\AA.  The data was reduced using KCWI's reduction pipeline (Version 1.1.0) and in-frame sky subtraction.

We optimally combined four exposures of lengths 1200s, 600s, 300s, and 100s to balance obtaining enough signal in the outer portions of the galaxy with saturation in the centre of the galaxy. For each spaxel, where wavelength pixels in the 1200s exposure were saturated, we replaced them with the pixels for that wavelength in the next longest exposure that was not saturated. The majority of the final data cube is obtained from the longest exposures, but the emission lines in the centre of the galaxy were often saturated in all but the 100s exposure.  Using this method, the final cube contains as much detail and little saturation as possible, with final dimensions of $24~\times~67$ pixels in the spatial direction ($32.6\arcsec~\times~19.5\arcsec$ or $12.7~\rm{kpc}\times7.6~\rm{kpc}$).

The continuum was fit in every spaxel with pPXF \citep{cappellari2017ppxf} using BPASS templates \citep[Version 2.2.1,][]{stanway2018reevaluatingbpass}.  We used BPASS templates including binary systems, with a broken powerlaw initial mass function with a slope of $-1.3$ between 0.1~M$_\odot$ and 1.0~M$_\odot$, a slope of $-2.35$ above  1.0~M$_\odot$, and an upper limit of 300~M$_\odot$.  The reddening of the galaxy spectra caused by Milky Way foreground extinction was included in the pPXF fitting and was determined using the \citet{calzetti2000dustcontent} extinction curve.

\section{Spatially Resolved Measurement of Outflows: \textsc{koffee}}
\label{sec:koffee}

\begin{figure}
    \centering
    \includegraphics[width=\columnwidth]{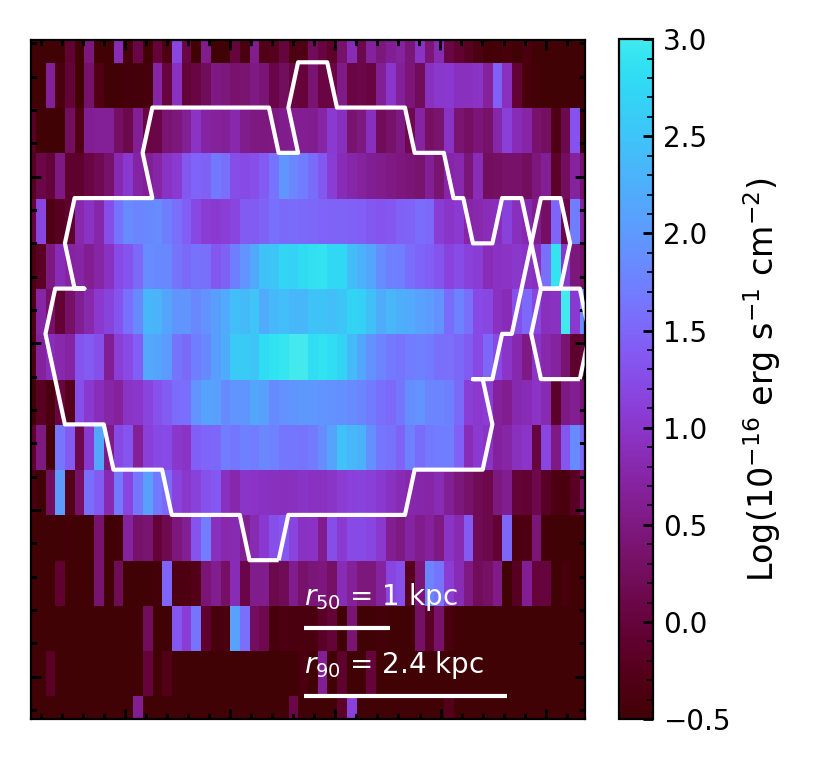}
    \caption{Map of the [OIII]~5007~\AA\ emission line flux from our KCWI data.  The white contours show where we have spaxels with a signal-to-noise of greater than 20 per pixel in a 5~\AA-wide continuum region blueward of the [OIII]~5007~\AA\ emission line.}
    \label{fig:OIII_map}
\end{figure}

\begin{figure*}
    \centering
    \includegraphics[width=\textwidth]{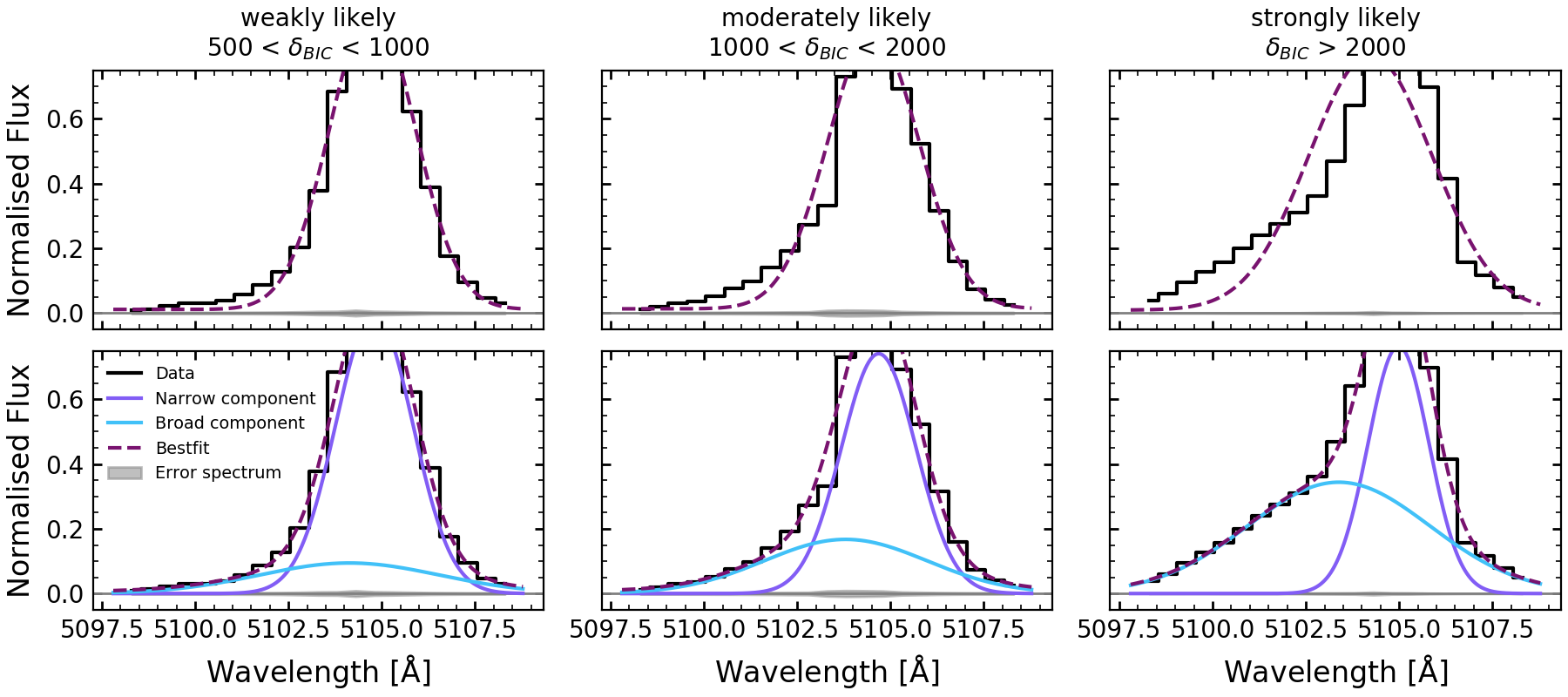}
    \caption{Three examples of fits of the continuum-subtracted [OIII]~$\lambda5007$ emission line made using \textsc{koffee}. The same spaxel is shown in each column with the fit made using a single Gaussian (\textit{top row}) and the fit made using two Gaussians (\textit{bottom row}).  Data and fits have been normalised to the peak.  The grey shaded area indicates the $1\sigma$ error spectrum.  The significance level of each double Gaussian fit corresponds to the amount by which the BIC$_{\rm{double~Gaussian}}$ is better than the BIC$_{\rm{single~Gaussian}}$ (see text).}
    \label{fig:example_fits}
\end{figure*}

In the case of resolved outflows we do not know {\em a priori} which spaxels require multiple Gaussian components and which do not. The simple assumption that all spaxels have an extra Gaussian, and thus an outflow, may lead to specious results. In this work we generate software, \textsc{koffee}\footnote{Our code, called \textsc{koffee} (Keck Outflow Fitter For Emission linEs), can be found here: \url{https://github.com/bronreichardtchu/koffee/tree/PaperI-code}.}, that applies a series of tests to each spaxel to determine if an outflow component is justified. 

\subsection{Overview of fitting method}
We only fit spaxels with a signal-to-noise (S/N) of greater than 20 per pixel in the continuum. This is measured in a region of bandwidth that is $\sim$5~\AA\ wide bluewards of the [OIII]~5007~\AA\ emission line. 
In Figure~\ref{fig:OIII_map} we show the [OIII]~5007~\AA\ emission line flux across the KCWI cube.  The white contours show where we have spaxels with a S/N~$>20$. 
In our observations of IRAS08 we find S/N~$>20$ to galactocentric radii beyond the 90\% radius of the optical continuum (6.1\arcsec or 2.4~kpc), and therefore are able to represent the outflow activity across the full face of the disk. 

An image of the [OIII] 5007~\AA\ line is shown in Fig.~\ref{fig:OIII_map}, with a white contour representing the area in which the continuum exceeds our threshold. The high S/N and spatial resolution of our data is sufficient to identify the star-forming ring as well as other areas of high star formation in the disk.  The [OIII]~5007 flux covers a difference of almost 3 orders of magnitude.  Typical flux values in the ring are $\sim$250-1000~$\times10^{-16}~\mathrm{erg~s}^{-1}~\mathrm{cm}^2$, where as in the fainter parts of the disk values can be as low as $\sim$3-10~$\times10^{-16}~\mathrm{erg~s}^{-1}~\mathrm{cm}^2$.  Under the typical assumption that outflow properties are linked to star formation and the emission line flux traces star formation to first order, the large range in flux values will allow us to detect a range of outflow behaviours.
We leave investigation of the impact on our results of varying this S/N cut to future work.  

For each spaxel with S/N~$>20$, \textsc{koffee} takes a continuum-subtracted emission line, and uses the python package \texttt{lmfit} \citep{newville2019lmfit0.9.14} using the default Levenberg-Marquardt least squares method to fit the line twice.  First with a single Gaussian,    
\begin{equation}
    f_{\lambda} = A \ exp\left[-\frac{(\lambda-\lambda_0)^2}{2\sigma^2}\right] + c,
\end{equation}
secondly with two Gaussians,

\begin{equation}
\begin{aligned}
    f_{\lambda} = & A_{\rm broad}\ exp\left [-\frac{(\lambda-\lambda_{\rm 0,broad})^2}{2\sigma_{\rm broad}^2}\right ]  
        +  \\ 
    &  A_{\rm narrow}\  exp[-\frac{(\lambda-\lambda_{\rm 0,narrow})^2}{2\sigma_{\rm narrow}^2}] + c,
\end{aligned}
\end{equation}
\vskip 5pt
\noindent where $A$ is the amplitude, $\lambda_0$ is the central wavelength, $\sigma$ is the standard deviation of the Gaussian, and $c$ is a constant used to account for any continuum remaining after the continuum subtraction.  The least squares fit is weighted by the error for each spectrum. 

\begin{figure*}
    \centering
    \includegraphics[width=\textwidth]{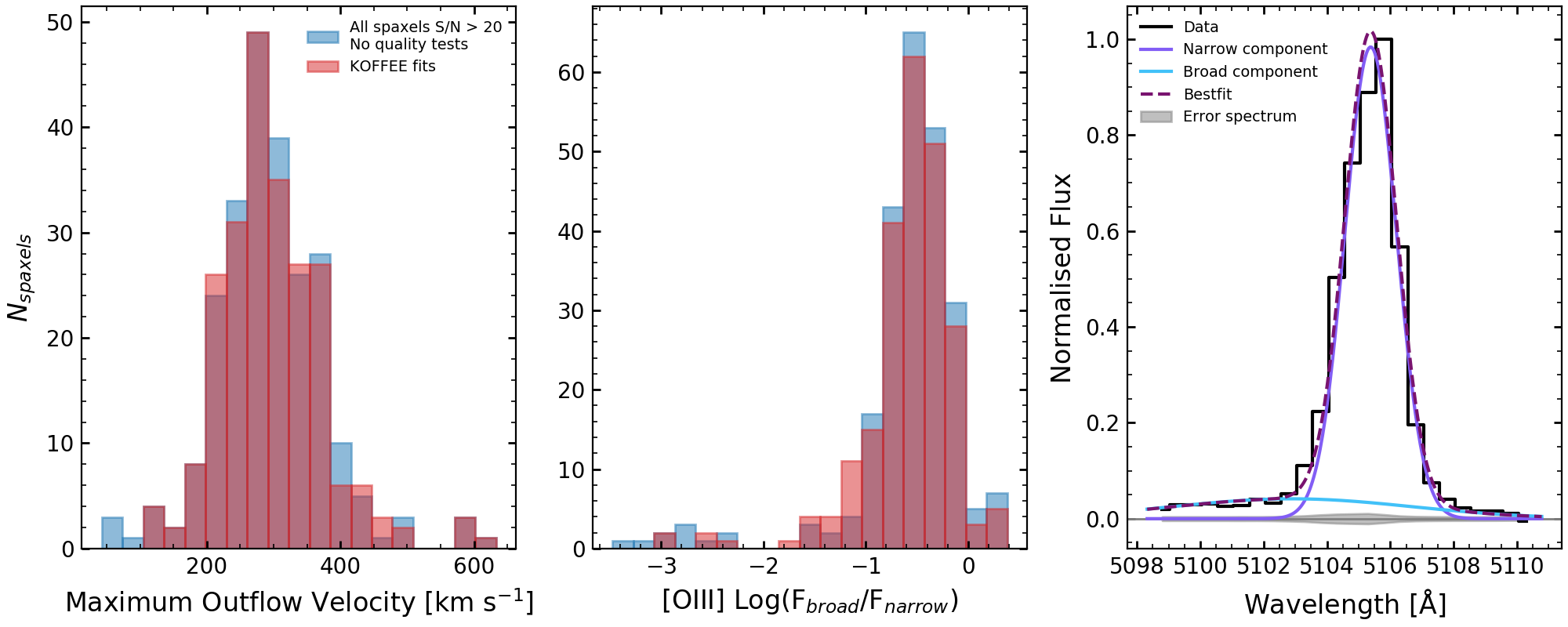}
    \caption{\textit{Left} and \textit{middle} panels show the distribution of outflow velocity and [OIII] flux ratio respectively for double Gaussian fitted spaxels.  In blue are the results if all 240 spaxels with a signal-to-noise ratio~$>20$ in the continuum are assumed to require a broad component.  In red are the results for the 230 spaxels which require a double Gaussian fit after undergoing the BIC selection test and the $\chi^2_{\rm blue}$ test required by our code (see Eq. \ref{eq:bic} and \ref{eq:bic_condition}). \textit{Right} is an example of a fit to the [OIII]~$\lambda5007$ emission line for a spaxel which does not require a second Gaussian in the fit, but is assumed to have one if our tests are not applied.  Fits such as this  over-estimating the number of spaxels where outflows are observable, and may increase scatter in low flux ratio parameter space.}
    \label{fig:all_vs_koffee_fits}
\end{figure*}

\subsection{Fitting Constraints}
The following constraints have been placed on the fitted Gaussians:  all Gaussians are required to have $\sigma>0.8$\AA\ ($\sim$47~km~s$^{-1}$).  This value is chosen as the minimum dispersion observable with KCWI for our settings.  To reduce the number of operations, $\sigma$ has not been convolved with the spectral resolution since the narrow line is unresolved and the broad lines are greater than the instrumental dispersion.  All Gaussians are constrained to have $A>0$.  

For the double Gaussian fit, the wavelength location of the emission line peak is used to define the expected $\lambda_{\rm 0,narrow}$.  From a visual inspection of our data for IRAS08, we find asymmetric emission lines skewed predominately towards the blue (e.g. Fig. \ref{fig:example_fits}).  The initial guess for $\lambda_{\rm 0,broad}$ is therefore set 0.1\AA\ ($\sim$6~km~s$^{-1}$) bluewards of $\lambda_{\rm 0,narrow}$.  $\lambda_{\rm 0,broad}$ is required to be within 5\AA\ ($\sim$294~km~s$^{-1}$) of $\lambda_{\rm 0,narrow}$.  Initially guesses for $A_{\rm narrow}$ and $A_{\rm broad}$ are given values of 70\% and 30\% of the emission line peak respectively.  The broad Gaussian is required to have $A_{\rm broad}<A_{\rm narrow}$.  
The initial guess for $\sigma_{\rm narrow}$ is 1.0\AA\ ($\sim$59~km~s$^{-1}$), and for $\sigma_{\rm broad}$ is 3.5\AA\ ($\sim$206~km~s$^{-1}$).  These initial values were chosen by trial and error. 

\subsection{BIC Test}
We use the Bayesian Information Criterion (BIC) to determine which spaxels are better fit by a double Gaussian.  The definition of the BIC is: 
\begin{equation}
    \mathrm{BIC} = \chi^2 + N_\mathrm{variables} \ln(N)  
     \label{eq:bic}
\end{equation}
where $N$ is the number of spectral data points, $N_{\rm variables}$ is the number of variables in the fit (4 for single Gaussians, 7 for double Gaussians), and $\chi^2=\sum(f_{\rm{data}}-f_{\rm{model}})^{2}/\sigma^2$ is the chi-square statistic calculated from the residuals of the Gaussian fits weighted by the data uncertainty, $\sigma$.  
A lower 
BIC value justifies the use of more model parameters to describe data.  The BIC provides a quantifiable and automated method of identifying those spaxels that require an outflow component.   We note that \texttt{lmfit}'s built-in function for the BIC does not follow this definition, and care should be taken before using their function for fitting purposes.

We define the difference in BIC values, 
\begin{equation}
    \delta_{BIC} = \mathrm{BIC}_{\rm{single~Gaussian}} - \mathrm{BIC}_{\rm{double~Gaussian}}.
     \label{eq:bic_condition}
\end{equation}
Typically $\delta_{BIC}=10$ is used to provide evidence against the fit with fewer parameters \citep{Kass1995BayesFactors, swinbank2019energetics, avery2021incidence}. In this paper We use a more stringent requirement and investigate the impact of this choice of $\delta_{BIC}$ on our results. We first consider all spaxels with $\delta_{BIC}<500$ to not have outflows that are detectable in our data. We then categorise the remaining spaxels, with outflows, according to their respective value of $\delta_{BIC}$ such that: \textit{strongly likely}, \textit{moderately likely} or \textit{weakly likely} to contain an outflow, corresponds to a BIC where $\delta_{BIC}>$~2000, 1000 and 500 respectively.  Example fits for each category are given in Fig. \ref{fig:example_fits}.  These category boundaries have been chosen after visual inspection of the fits. We note that we find that the minimum $\delta_{BIC}$ is significantly higher than the conventional value described in statistical texts. This is likely due to the known result that even in more typical spiral galaxies emission lines have been shown to be better fit by multiple Gaussian components \citep{ho2014sami}. Moroever, imperfect continuum subtraction is a reality in young stellar populations, and in some cases the software may mistake this for a low-flux broad component (e.g. Fig.~\ref{fig:all_vs_koffee_fits}).  We, therefore, choose the larger value in order to isolate the broad component that we are assuming to be representative of the outflow. We suggest that before implementing the BIC as an automated method of decision-making in fitting galaxy emission lines, the chosen cutoffs should be rigorously investigated rather than applying values that are not specifically designed for isolating components of galaxy spectra.

We point out those spaxels in the  \textit{weakly likely} category of Fig.~\ref{fig:example_fits}. According to standard practice these fits have ``strong evidence" from their BIC values to be considered outflow spaxels, however many of these spaxels may not be judged as needing an extra component in by-eye analysis.  

\subsection{\texorpdfstring{$\chi^2_{\rm blue}$}{TEXT} Test}
We perform a second test 
intended to specifically determine if the region of the spectrum that may be dominated by outflow is well fit by the model chosen after the BIC test. We calculate $\chi_{\rm blue}^2\equiv\Sigma_i (f_{\rm model} - f_{\rm data})^{2}/\sigma_i^2$ in a 4.0\AA\ region blue-ward of  $\lambda_{\rm 0,narrow}-\sigma_{\rm narrow}$ for all spaxels using the best fitting model. If  $\chi_{\rm blue}^2~>~1.5$, the emission line is refit with a double Gaussian fit using new initial guesses: $\lambda_{\rm 0,broad}$ is shifted 4.0\AA\ ($\sim$240~km~s$^{-1}$) to shorter wavelength from $\lambda_{\rm 0,narrow}$, and $\sigma_{\rm broad}$ is increased to 8.0\AA\ ($\sim$480~km~s$^{-1}$).  These new initial guesses force the software to explore parameter space including broader outflows with a greater mean offset from the narrow Gaussian.  The resulting double Gaussian fit is adopted if it has improved the calculated $\chi_{\rm blue}^2$ and has a lower BIC value than the single Gaussian fit.

\subsection{Fitting [OIII] and \texorpdfstring{H$\beta$}{TEXT}}
\textsc{koffee} fits the [OIII]~$\lambda5007$ emission line first, using the results to inform the fit for the H$\beta$ emission line. We do this for multiple reasons. 
There is a small diminution in the H$\beta$ line due to absorption, which has been removed in the continuum subtraction (see Sec. \ref{sec:observations}), and this may bias results. Moreover,  H$\beta$ is significantly fainter than [OIII] and the S/N is likely to lead to less trustworthy measurements at low surface brightness. 

We first assume that the BIC and $\chi^2_{\rm blue}$ results on the [OIII]~$\lambda5007$ line apply to H$\beta$. The initial parameter guesses for $\sigma_{\rm narrow}$, $\sigma_{\rm broad}$ and $\lambda_{\rm 0,narrow}-\lambda_{\rm 0,broad}$ are given by the fitted parameters from the [OIII]~$\lambda5007$ results.  The parameters are allowed to vary by 1.5\AA\ ($\sim 90~\mathrm{km}~\mathrm{s}^{-1}$) from those found for [OIII]~$\lambda5007$.

The $\chi^2_{\rm blue}$ test is then performed on the double Gaussian H$\beta$ fits.  If $\chi^2_{\rm blue}>1.5$ for H$\beta$, the emission line is refit with only 0.5\AA\ ($\sim 30~\mathrm{km}~\mathrm{s}^{-1}$) variation allowed for both parameters. Due to the lower S/N on H$\beta$, we base our refits more heavily on the [OIII] parameters.  The resulting double Gaussian fit is adopted if it has improved the calculated $\chi_{\rm blue}^2$ and has a lower BIC value than the single Gaussian fit. If $A_{\rm broad}/A_{\rm narrow}>0.9$ we perform a BIC test to decide if the second Gaussian is warranted. If $A_{\rm broad}/A_{\rm narrow}>0.98$ then we assume that only a single Gaussian is observable in H$\beta$. We find that the outflow velocities derived from H$\beta$ are of order 10\% lower than those from [OIII].  


\subsection{Impact of fitting tests on results}

In Fig.~\ref{fig:all_vs_koffee_fits} we show the impact of these tests on our main results for derived outflow properties, comparing the results from \textsc{koffee} and those making the assumption that all spaxels require a double Gaussian fit. Overall, the biases in IRAS08 are not strong, though this may be due to the widespread nature of outflows in this galaxy. Galaxies that are less covered by outflows may have more biases.  
If we only perform the BIC selection test, 199 out of all 240 spaxels require the double Gaussian fit. 
When we perform both the BIC selection test and the $\chi^2_{\rm blue}$ test, 230 spaxels require the double Gaussian fit.  

When all spaxels are assumed to require a double Gaussian fit, there is a population that have low flux ratios (middle panel of Fig.~\ref{fig:all_vs_koffee_fits}). 
Many of these spaxels do not truly require a second Gaussian to fit the [OIII]~$\lambda5007$ line. An example of such a fit is given in the right-most panel of Fig. \ref{fig:all_vs_koffee_fits}. 


\section{Results} \label{sec:results}

\subsection{Outflow Velocity}
\begin{figure*}
    \centering
    \includegraphics[width=\linewidth]{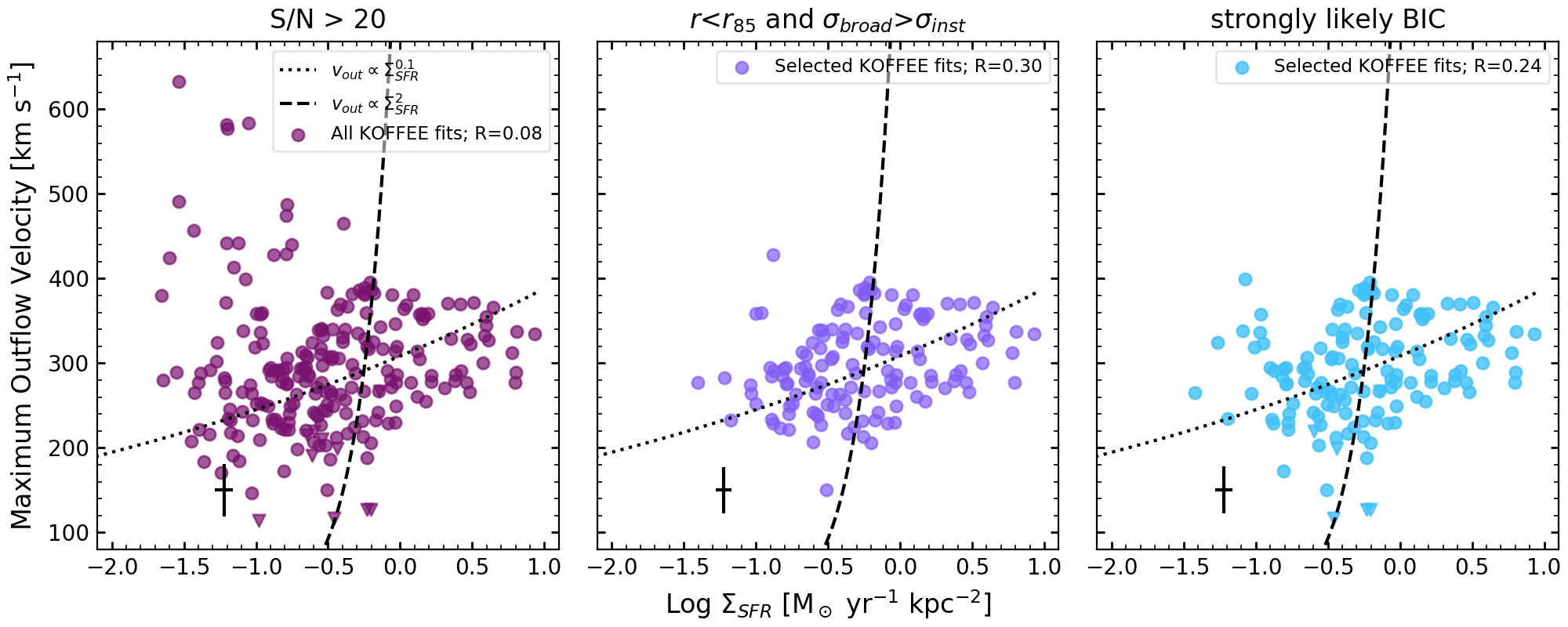}
    \caption{The maximum outflow velocity, \vout\, is plotted against the star formation rate surface density, \sigmasfr, for \textit{left:} all KCWI spaxels in IRAS08 which require a double Gaussian fit in our data; \textit{middle:} spaxels requiring a double Gaussian fit which are within $r_{85}$ and have a fitted $\sigma_{\rm broad}$ greater than the instrument dispersion $\sigma_{\rm inst}$; and \textit{right:} spaxels which are fit with a \textit{strongly likely} outflow component according to our BIC test.  
    The Pearson's correlation coefficient, R, is given in the legend of each panel.
    Two models for the relationship between \vout\ and \sigmasfr\ have also been plotted.  The dashed line represents a momentum-driven outflow, where the outflow is driven by radiative pressure from young stars \citep[\vout$\propto$\sigmasfr$^2$;][]{murray2011radiationpressure}.  The dotted line represents an energy-driven outflow, where the outflow is driven by supernova feedback \citep[\vout$\propto$\sigmasfr$^{0.1}$;][]{chen2010absorption}.  Our data most closely resembles the trend expected of an energy-driven outflow.}
    \label{fig:sigma_sfr_vel_max}
\end{figure*}

The relationship between outflow kinematics and the properties of the region launching them has historically been used to discriminate between physical models of the launching mechanism \citep[e.g.][]{chen2010absorption,murray2011radiationpressure,newman2012sins}. We, therefore, use our results derived from \textsc{koffee} to study such correlations. We find that in IRAS08 this relationship is more consistent with shallow slopes, similar to energy driven winds. 

\vskip 5pt
In Figure \ref{fig:sigma_sfr_vel_max} we show the \sigmasfr\ plotted against the maximum outflow velocity, \vout\, defined as
\begin{equation}
    v_{\rm out} = |v_{\rm narrow}-v_{\rm broad}| + 2\sigma_{\rm broad}
    \label{eq:outflow_vel}
\end{equation}
where $v_{\rm narrow}$ and $v_{\rm broad}$ are the velocities at the centre of the narrow and broad Gaussians respectively, and $\sigma_{\rm broad}$ is the standard deviation of the broad Gaussian for the [OIII]~$\lambda5007$ fits which has been corrected for the instrumental velocity dispersion of 0.7~\AA\ or 41.9~km~s$^{-1}$.  In general, these quantities need to be corrected for galaxy inclination, however this is not important in our face-on galaxy.  Equation~\ref{eq:outflow_vel} is similar to outflow velocity measurements that have been used in the literature \citep[e.g.][]{genzel2011sins, davies2019kiloparsec}. We use this definition to make our velocities measured from emission lines more comparable with the maximum velocities measured from absorption line studies.  Each data point in Fig. \ref{fig:sigma_sfr_vel_max} represents a KCWI spaxel in which a double Gaussian fit is required.  Spaxel sizes are of order $\sim100\times500$~pc. In IRAS08 we find that the average spaxel has a median \vout\ of 288~km~s$^{-1}$ with root-mean-square scatter of 79~km~s$^{-1}$. The typical measurement uncertainty on a single spaxel is of order 10-30~km~s$^{-1}$. 

To calculate the \sigmasfr\ for each spaxel, we use
\begin{equation}
    SFR = \mathrm{C}_{\mathrm{H}\alpha} \frac{L_{\mathrm{H}\alpha}}{L_{\mathrm{H}\beta}} 10^{-0.4A_{\mathrm{H}\beta}} L_{\mathrm{H}\beta}
    \label{eq:sfr}
\end{equation}
and divide by the KCWI spaxel size.  Here C$_{\mathrm{H}\alpha}=10^{-41.257}$ is the scale parameter \citep{hao2011dustcorrected}, $L_{\mathrm{H}\alpha}/L_{\mathrm{H}\beta}=2.87$ is the luminosity ratio \citep{calzetti2001dustopacity}, $A_{\mathrm{H}\beta}$ is the extinction and $L_{\mathrm{H}\beta}$ is the observed H$\beta$ luminosity.  We have used the observed emission line ratio H$\beta/$H$\gamma$ to correct for extinction in the emission lines \citep{cardelli1989relationship, calzetti2001dustopacity}.  To calculate $L_{\mathrm{H}\beta}$ for each spaxel we have used the narrow Gaussian fit to H$\beta$ from \textsc{koffee}.  The flux contributed by the broad component is interpreted to be outflowing gas, and so is left out of the calculation.

Typical studies do not remove the outflow component from their calculations of the \sigmasfr\ from ionised gas emission. 
Using the combined broad+narrow flux we obtain a 
total SFR similar to typical estimates \citep[e.g.][]{ostlin2009lymanalphamorphology}.  
Averaging over all spaxels, we find that taking the outflow into account by removing it causes an average decrease in \sigmasfr\ of $\sim$25\%. This implies that in star-forming galaxies, such as those at $z>1$, outflows only contribute to a systematic uncertainty in SFR of order $0.15$~dex. This may increase for galaxies with higher \sigmasfr\ than our target.

In Fig.~\ref{fig:sigma_sfr_vel_max} we also show two popularly used models for the relationship between \vout\ and \sigmasfr.  The dashed line shows the expected relationship if the outflows are momentum-driven 
\citep[\vout$\propto$\sigmasfr$^2$;][]{murray2011radiationpressure}.  The dotted line shows the expected relationship if the outflows are energy-driven 
\citep[\vout$\propto$\sigmasfr$^{0.1}$;][]{chen2010absorption}.

As discussed above, we do not know {\em a priori} which spaxels are appropriate for comparison of outflows to physical models.  In Fig.~\ref{fig:sigma_sfr_vel_max} we, therefore, make three separate assumptions on which data is appropriate to compare to the models: all data points with S/N~$>20$ in the continuum and at least \textit{weakly likely} double Gaussian fits (left); those data points within the 85\% radius ($r_{85}\sim4.4$\arcsec or 1.7~kpc) and which have resolved dispersions (middle); and those spaxels which have \textit{strongly likely} double Gaussian fits (see Sec. \ref{sec:koffee}). 

In all three panels of Fig.~\ref{fig:sigma_sfr_vel_max} we find that for \sigmasfr~$>0.1~$\sigmasfrunits\ there is a positive correlation between \sigmasfr\ and \vout. For both the physically selected panel (middle) and the data points selected by stronger BIC fits (right) we find 
a strong correlation, with Pearson's coefficients of 0.30 and 0.24 respectively (see Table~\ref{tab:sigma_sfr_vel_max}). Moreover, the slope of this correlation is shallow, and thus consistent with typical expectations from energy-driven wind models, where \sigmasfr~$\propto v_{\rm out}^{0.1-0.2}$ \citep{chen2010absorption, li2017supernovaedriven}. 

In the left-hand panel of Fig.~\ref{fig:sigma_sfr_vel_max}, 
there is an increase in the typical value of \vout\ below a \sigmasfr\ of  $\sim0.1~$\sigmasfrunits. We find that this increase is due to broad components located at large radius, $r>4.4\arcsec$ ($r_{85}$). 
The middle panel shows the spaxels remaining after we exclude the 36\% of our double Gaussian fitted spaxels located beyond $r_{85}$. We note that 4\% of our double Gaussian fitted spaxels have broad Gaussian components that are equivalent to the instrumental dispersion of KCWI for our settings. These are marked as triangles in the left-hand panel.   The remaining spaxels (62\% of the total) have a median \vout\ of 293~km~s$^{-1}$ with root-mean-square scatter of 51~km~s$^{-1}$. Restricting the spaxels included 
in this way shows that the upturn below $0.1~$\sigmasfrunits\ is dominated by spaxels at large radius.  
We note that, as discussed above, IRAS08 is experiencing a distant interaction with a smaller galaxy, and so some mechanism other than star-formation feedback may be driving the kinematics at the edge of the galaxy disk.  Excluding high radius points, our data is consistent with the trend expected from the energy-driven model. 


\begin{table*}
    \centering
    \caption{Statistics for the relationship between \sigmasfr\ and \vout\ when considering all of the spaxels with a S/N~$>20$ (top section of table, see Fig.~\ref{fig:sigma_sfr_vel_max}), only spaxels with S/N~$>20$ and \sigmasfr~$>0.1~$\sigmasfrunits\ (middle section of table), and spaxels with a S/N~$>20$ when data is binned to circularised diameters of $\sim600$~pc (bottom section of table, see Section~\ref{sec:impact_spatial_res} and Fig.~\ref{fig:sig_sfr_out_vel_binned}).  Columns are for the categories defined in the panels of Fig. \ref{fig:sigma_sfr_vel_max}. $N$ is the number of spaxels in each category. R is the Pearson's Correlation Coefficient, and the p-value gives the probability that a similar R-value could be given by uncorrelated systems for all spaxels. 
    }
    \label{tab:sigma_sfr_vel_max}
    \begin{tabular}{lccc}
         & S/N~$>20$ and $\delta_{\rm BIC}>500$ & $\delta_{\rm BIC}>500$, $r<r_{85}$ and $\sigma_{\rm broad}>\sigma_{\rm inst}$ & Strongly likely BIC $\delta_{\rm BIC}>2000$ \\
        \hline 
        \hline 
        $N$ & 230 & 137 & 140\\
        R & 0.08 & 0.30 & 0.24\\ 
        p-value & 0.24 & 4x10$^{-4}$ & 2x10$^{-3}$\\ 
        
        \hline
        \multicolumn{4}{c}{\sigmasfr~$>0.1~$\sigmasfrunits} \\
        \hline 
        $N$ & 186 & 132 & 133\\ 
        R & 0.23 & 0.28  & 0.30\\ 
        p-value & 2x10$^{-3}$ & 1x10$^{-3}$ & 5x10$^{-4}$\\ 
        
        \hline
        \multicolumn{4}{c}{Circularised Diameter 0.6~kpc Bins} \\
        \hline 
        $N$ & 89 & 32 & 74\\ 
        R & 0.13 & 0.51 & 0.17\\ 
        p-value & 0.21 & 0.003 & 0.14\\ 
         
    \end{tabular}
\end{table*}


In the right-hand panel of Fig.~\ref{fig:sigma_sfr_vel_max}, we have restricted our sample to include only spaxels in which are \textit{strongly likely} to contain an outflow according to our BIC test.  This includes 61\% of the total spaxels which \textsc{koffee} originally fit with double Gaussians. These spaxels have a median \vout\ of 291~km~s$^{-1}$ with root-mean-square scatter of 58~km~s$^{-1}$.  The scatter which we find here is on par with the velocity resolution, which may be a contributing factor to the distribution of velocities.

There is an often quoted fiducial threshold value of $0.1~$\sigmasfrunits\ for entire galaxy measurements, below which outflows are generally not observed \citep[e.g.][]{heckman2002galacticsuperwinds, veilleux2005galacticwinds, heckman2015systematic}.  
We therefore compare the \sigmasfr$-$\vout\ relationship for all spaxels, and spaxels with \sigmasfr~$>0.1~$\sigmasfrunits. Results are given in the top section of Table~\ref{tab:sigma_sfr_vel_max} for all spaxels and the middle section of Table~\ref{tab:sigma_sfr_vel_max} for spaxels with \sigmasfr~$>0.1~$\sigmasfrunits. 
The correlation between \sigmasfr\ and \vout\ is strongest for the case in which all \sigmasfr\ with physically selected or \textit{strongly likely} BIC values are included, as indicated by the correlation coefficients. 
Moreover, as is clear in the right two panels of Fig.~\ref{fig:sigma_sfr_vel_max} there is no evident break in the correlation at this proposed threshold value. 

Conversely for more vigorous star formation, \cite{newman2012sins} finds ``strong" outflows are restricted to \sigmasfr~$>1~$\sigmasfrunits. We find that outflows are widespread with values reaching \sigmasfr\ multiple orders of magnitude below this value. We will return to this in a subsequent discussion of mass-loading factor, and also in the Discussion we address the impact on estimates of the covering fraction of outflows in star forming disks.


\subsection{Impact of Region Size Averaging on the \texorpdfstring{\sigmasfr~$-v_{\rm out}$}{TEXT} Relationship}
\label{sec:impact_spatial_res}

We investigate the effect of increasing the spatial sampling scale (i.e.\ resolution) on the relationship shown in Fig.~\ref{fig:sigma_sfr_vel_max}. We note that our intent is to understand the impact of physical sampling scale on the correlation of  \vout\ and \sigmasfr, which is subtly different than the blurring due to observed resolution on the sky. Our procedure is similar to what was carried out by \cite{davies2019kiloparsec} on stacks of regions from different galaxies at larger redshift. We intend a study of spatial resolution in a follow up paper that will include a sample larger than one galaxy.

\begin{figure*}
    \centering
    \includegraphics[width=\textwidth]{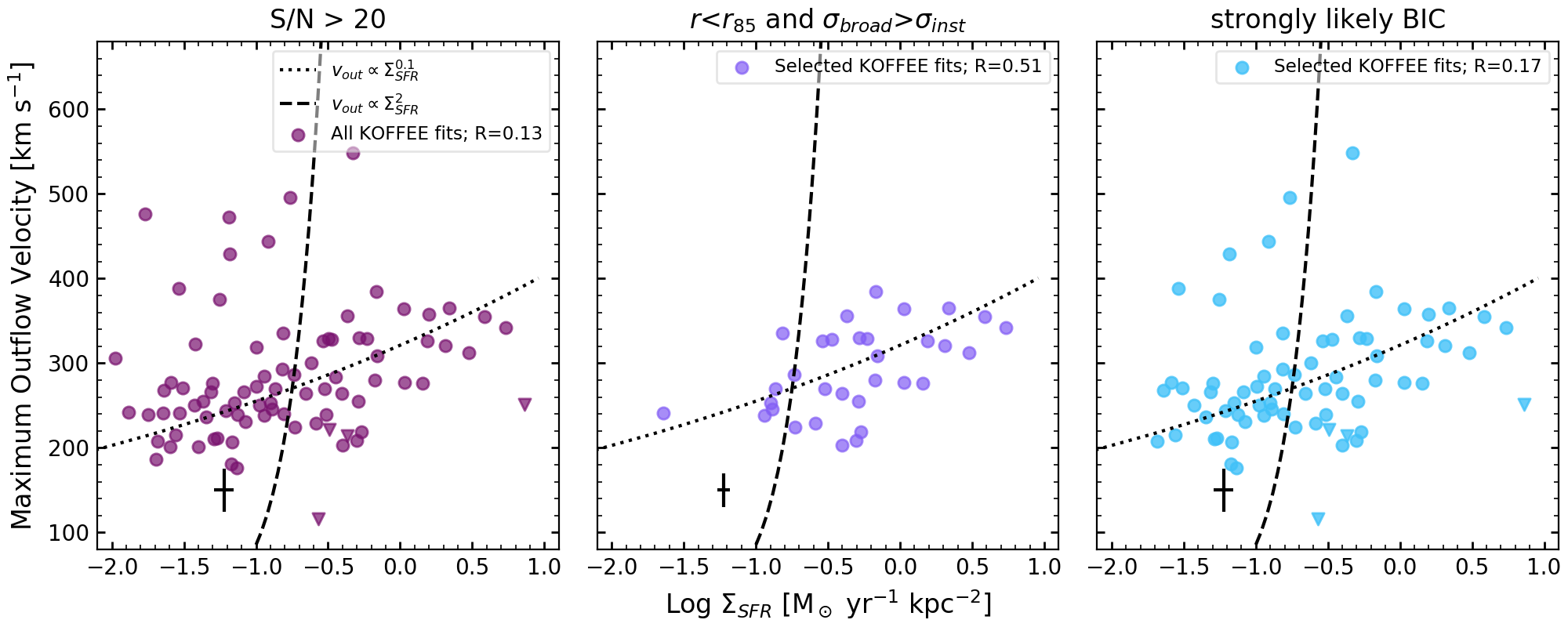}
    \caption{The same as for Figure~\ref{fig:sigma_sfr_vel_max}, with our KCWI spaxels binned 5x1 to give 1.45\arcsec x1.36\arcsec\ bins.  For IRAS08 at a distance of $\sim$82~Mpc, these are 0.56~kpc~x~0.53~kpc bins with an area of 0.3~kpc$^2$ and a circularised diameter of $\sim600$~pc. Increasing the physical size of the bin in which \sigmasfr\ and \vout\ are measured clearly creates stronger correlations between these parameters.}
    \label{fig:sig_sfr_out_vel_binned}
\end{figure*}

\begin{figure}
    \centering
    \includegraphics[width=\columnwidth]{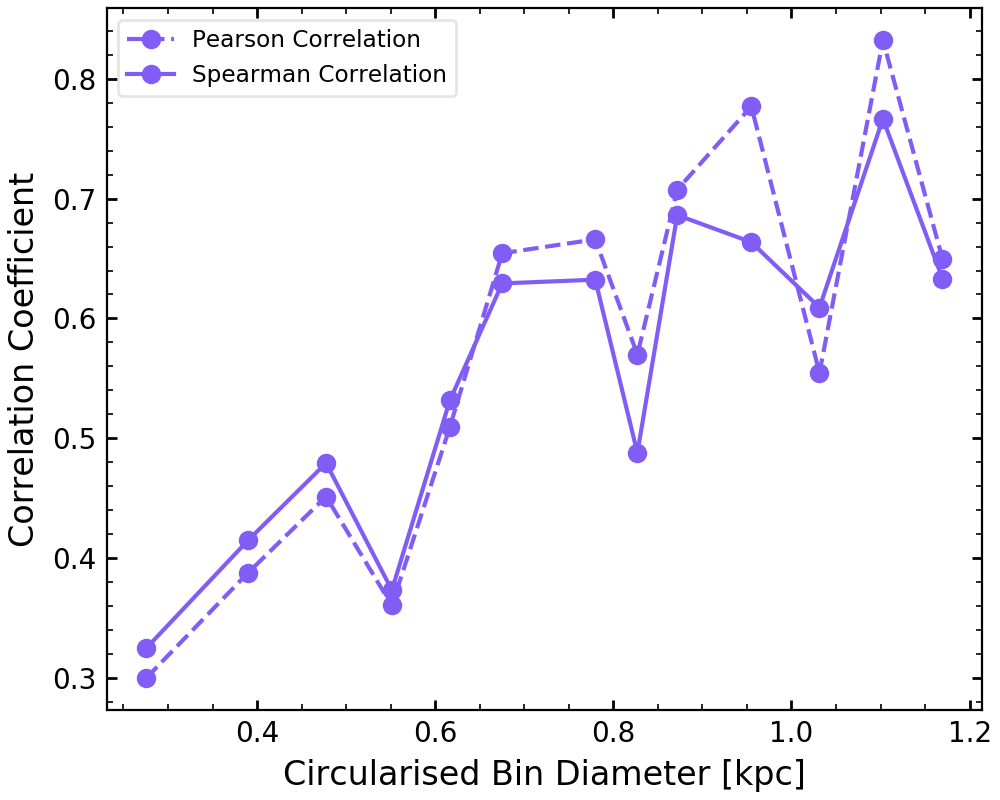}
    \caption{The Pearson (dashed line) and Spearman (solid line) correlation coefficients between the log of the outflow velocity $\log_{10}$(\vout) and the log of the star formation rate surface density $\log_{10}$(\sigmasfr) are plotted against the circularised bin diameter.  For clarity only the correlation coefficients for the physically motivated selection of points (middle panels of Fig.~\ref{fig:sigma_sfr_vel_max} and Fig.~\ref{fig:sig_sfr_out_vel_binned}) are shown.} 
    \label{fig:bin_correlations}
\end{figure}

We binned our spaxels to 13 different region sizes, increasing the sampling iteratively and then refit the relationship. This begins with 0.11$\times$0.53~kpc$^{2}$ (1$\times$1 binning) and increases to  $\sim$1.0$\times$1.1~kpc$^2$ (9$\times$2 binning).  The smallest bin size is the size of a single KCWI spaxel and is designed to sample half the nominal seeing at Keck ($\sim$0.7\arcsec or $\sim$0.27~kpc). We represent these rectangular bins on the x-axis of Fig.~\ref{fig:bin_correlations} by the equivalent circularised diameter for each binning scale.  Emission lines were shifted to have the same central velocity before being combined to remove line-broadening caused by local variations in the systemic velocity. This shift is done to identify the physical resolution at which the relationship between outflows and galaxy become most correlated. The typical shift increases as the binning size is increased, and ranges on average from 2 km/s when binning 2x1 spaxels (0.2~kpc x 0.5~kpc) to 240~km~s$^{-1}$ when combining 9x2 spaxels (1~kpc $\times$ 1~kpc). 
Shifting the emission lines creates a systematic difference between our resolution study and observations of outflows at lower resolution \citep[e.g.][]{newman2012sins}. The typical spatial resolution (in FWHM) of $z\sim2$ outflow measurements with adaptive-optics data \citep[e.g.][]{newman2012sins,genzel2011sins} would still be larger than our largest circularised diameter in this study. The binned spaxels were then run through \textsc{koffee}, and \vout\ and \sigmasfr\ re-calculated for each bin. Figure~\ref{fig:sig_sfr_out_vel_binned} shows an example where we have binned the spaxels to a circularised diameter of $\sim600$~pc.  The statistics for this example are given in the bottom section of Table~\ref{tab:sigma_sfr_vel_max}.

In Fig.~\ref{fig:bin_correlations} we show how the Pearson and Spearman correlation coefficients for the physically motivated selected spaxels 
depend on circularised bin diameter.  
The Pearson correlation coefficient measures the linear correlation between two sets of data.  The Spearman correlation coefficient measures nonparametric correlation between two sets of data that are not necessarily normally distributed.  If the correlation is close to linear and there are no strong outliers, the Spearman correlation coefficient will be similar to the Pearson correlation coefficient.
We find that decreasing the spatial resolution-- or increasing the bin size-- corresponds with an increase in the correlation coefficients between \sigmasfr\ and \vout. 
This result can also be seen by comparing the Pearson's correlation coefficients reported in the top and bottom sections of Table~\ref{tab:sigma_sfr_vel_max}.

To explain the above result, we discuss a geometric as well as a timescale motivated argument. We note that these are not mutually exclusive, nor are they the only possible causes.

Using the \sigmasfr~$-$\vout\ correlation makes the implicit assumption that, at our resolution, the outflowing gas is colocated on the sky with the site of the star formation that launched it. However, gas moving at an angle could be observed above an area that is not the launch site, with an unrelated \sigmasfr. We have chosen a target that is near-to face-on.  However, the outflows themselves may not occur precisely perpendicular to the face of the disk.  This could increase scatter in the \sigmasfr~$-$\vout\ distribution.  As the bin size increases, the observed outflowing gas is more likely to be directly linked to the underlying observed \sigmasfr, increasing the correlation between points in the \sigmasfr~$-$\vout\ distribution. 

Secondly, during the time between when the outflow was launched and when we observe it, the underlying properties of the star forming region may have changed, causing the \sigmasfr\ measured within the bin to change.  
For example, the supernova driving the winds may mark a decrease in the emission line flux from the HII region over time. This could lead to high \vout\ points being co-located with lower than expected \sigmasfr. Increasing the bin size will then include neighbouring star formation regions. The outflow velocity may better correlate with the ``average" local \sigmasfr\ in the local region. 


It is unclear whether one of these is the dominant effect, or if a combination of the two is causing the increase in correlation between \vout\ and \sigmasfr. More work on larger samples of objects would be informative. 

\subsection{Mass Loading Factor}
\label{sec:flux_mlf}

\begin{figure*}
    \centering
    \includegraphics[width=\textwidth]{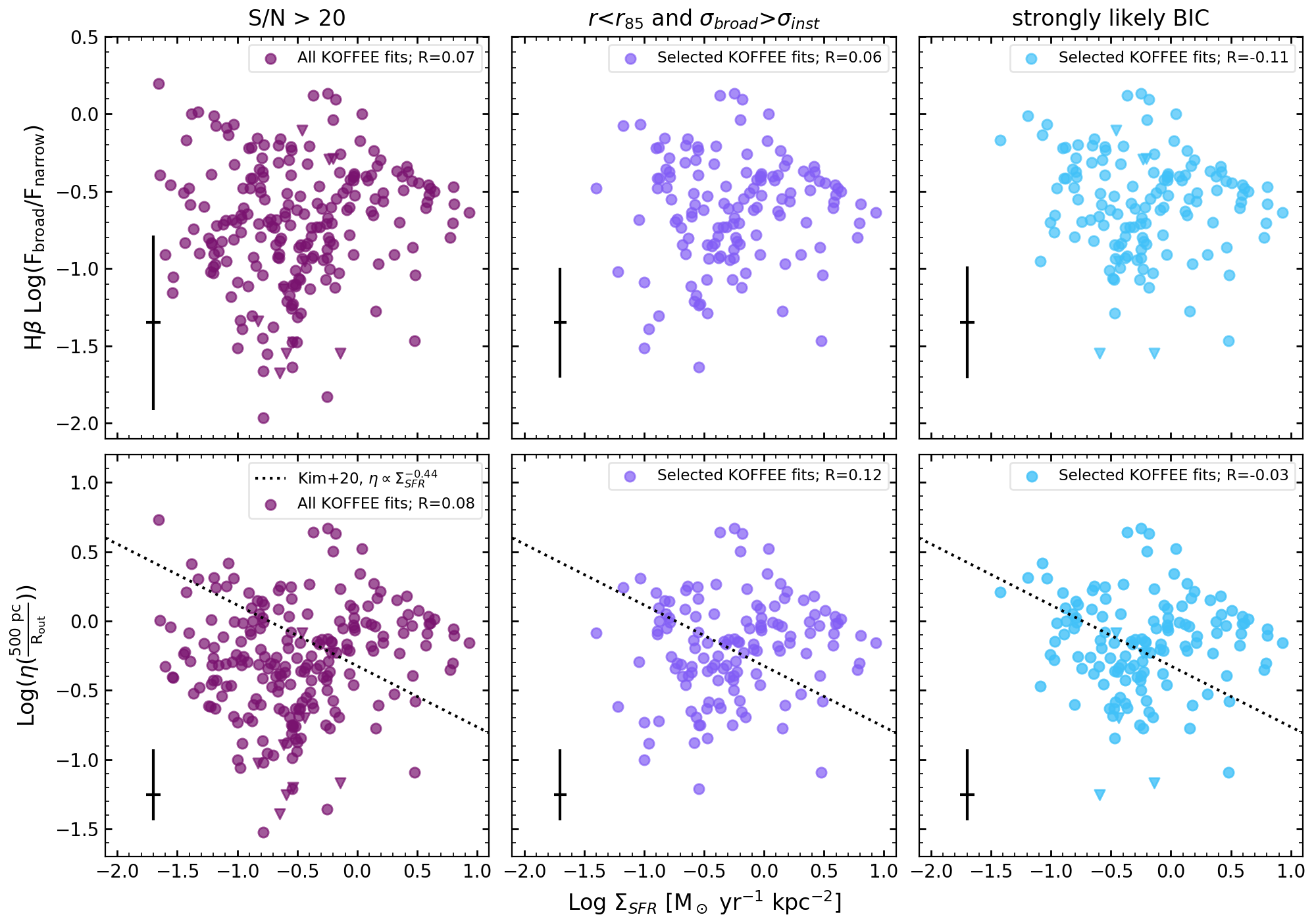}
    \caption{The log of the observed H$\beta$ broad-to-narrow flux ratio F$_{\rm broad}$/F$_{\rm narrow}$ (\textit{top row}), and the log of the mass loading factor $\eta$ (\textit{bottom row}) are plotted against star formation rate surface density, \protect\sigmasfr.  Panels are defined as in Fig. \ref{fig:sigma_sfr_vel_max}.  
    The Pearson's correlation coefficient, R, is given in the legend of each panel. 
    A model for the relationship between $\eta$ and \protect\sigmasfr\ has also been plotted in each panel of the bottom row (dotted line).  The model represents outflows driven by supernova feedback \citet{kim2020firstresultssmaug}, with an arbitrary scaling factor chosen to match our data. Our data follows a similar trend as the slope found by \citet{kim2020firstresultssmaug}.}
    \label{fig:mlf_hbeta_sig_sfr}
\end{figure*}

The mass loading factor is the ratio of the mass outflow rate to the star formation rate $\eta = \Dot{M}_{\rm out}/\mathrm{SFR}$. Similarly to the kinematics, the mass loading factor is commonly used to test the physical models driving the outflow.  It is also used to understand the rate of gas exiting the galaxy, or local region, due to feedback. There are a significant number of uncertainties in deriving $\eta$ which may vary within a galaxy. We consider these in this subsection. 

The mass outflow rate is defined as
\begin{equation}
    \Dot{M}_{\rm out} =  \frac{1.36m_\mathrm{H}}{\gamma_{\mathrm{H}\beta}n_e} \left(\frac{v_{\rm out}}{R_{\rm out}}\right) L_{\rm H\beta,broad}
    \label{eq:mass_outflow_rate}
\end{equation}
where $m_\mathrm{H}$ is the atomic mass of H, 
$\gamma_{\mathrm{H}\beta}$ is the H$\beta$ emissivity at $T_e=10^4$~K ($\gamma_{\mathrm{H}\beta}=1.24\times10^{-25}~\mathrm{erg~cm}^3~\mathrm{s}^{-1}$), $n_e$ is the local electron density in the outflow, \vout\ is the maximum outflow velocity which we found in the previous section, $R_{\rm out}$ is the radial extent of the outflow,  and $L_{\rm H\beta,broad}$ is the extinction-corrected H$\beta$ luminosity of the broad component.  Dividing the mass outflow rate by the SFR defined in Equation \ref{eq:sfr}, the mass loading factor is 
\begin{equation}
    \eta =  \frac{1.36m_\mathrm{H}}{\mathrm{C}_{\rm H\alpha}\gamma_{\rm H\beta}n_e} 10^{0.4A_{\rm H\beta}}\left(\frac{v_{\rm out}}{R_{\rm out}}\right) \frac{L_{\rm H\beta,broad}}{L_{\rm H\beta,narrow}}.
    \label{eq:mass_loading_factor}
\end{equation}

For the electron density in the outflow, we adopt the value of $n_e = 380~\mathrm{cm}^{-3}$, consistent with measurements by \cite{forsterschreiber2019kmos3d} and \cite{newman2012shocked}. The range in likely values of the electron density of outflows is of order $\sim$50-700$\mathrm{cm}^{-3}$, which translates to a systematic uncertainty of order $\sim$1~dex on $\eta$.

A significant source of systematic uncertainty in $\Dot{M}_{\rm out}$ is the radial extent of the outflow, $R_{\rm out}$. Along with observational measurement there are intrinsic physical uncertainties in assuming a single scale size for the outflowing gas in a region. The true value of $R_{\rm out}$ may vary point-to-point in a galaxy, and may also vary with time in a specific line-of-sight. For each individual spaxel, the  $\Dot{M}_{\rm out}$ may be both gas originating from launch sites colocated within that spaxel, and also tangentially launched gas from nearby spaxels. The choice of a single $R_{\rm out}$ for each spaxel therefore depends on our assumptions of the region contributing to the observed outflow.  

We have assumed an $R_{\rm out}$ of 500~pc for IRAS08 based on multiple lines of argument from nearby well-studied starbursts and our observations of IRAS08.  M82 and NGC~253 are very well-known, edge-on starbursting galaxies, in which the minor axis gas is found to have scale lengths of $400-700$~pc \citep{leroy2015multiphaseM82} and $200-500$~pc  \citep{krieger2019molecularoutflow, bolatto2013suppression} respectively. Similar results are found in NGC~1482 \citep{veilleux2002identificationNGC1482}. Alternatively, \cite{chisholm2016robust} use photoionisation modelling of absorption features to find that the majority of the outflowing mass from NGC~6090 is within 300~pc of the starburst. More work is direly needed on the extent of outflows in starbursting systems. Nonetheless the work to date suggests that scales of hundreds of parsecs are appropriate.  
Note due to this uncertainty, we express the mass-loading factor in units of 500~pc/$R_{\rm out}$ to emphasise that this assumption heavily affects this important quantity.


The errorbars for $\eta$ in Fig.~\ref{fig:mlf_hbeta_sig_sfr} include an estimate of the uncertainty in $R_{\rm out}$. For an upper limit, we assume that the maximum value of $R_{\rm out}$ is not larger than the 90\% radius of the galaxy (2.4~kpc). For the lower limit, we assume that the bulk of the outflowing material is observed close to its launching site. Simulation work often chooses the scale-height of the disk as a metric. This would be of order $100-200$~pc in a galaxy like IRAS08. For convenience, we use a minimum $R_{\rm out}$ of 350~pc, which is chosen to be similar to the resolution of our KCWI spaxels.  This reasoning is similar to that in \cite{davies2019kiloparsec}. The total range in viable assumptions for $R_{\rm out}$ implies a systematic uncertainty on the absolute value of $\eta$ of roughly an order-of-magnitude. 

In Figure \ref{fig:mlf_hbeta_sig_sfr} we plot the H$\beta$ broad-to-narrow flux ratio against \sigmasfr\ in the top row, and $\eta$ against \sigmasfr\ in the bottom row.  The columns represent the same assumptions as in Figure \ref{fig:sigma_sfr_vel_max}.  
In IRAS08 we find a total mass-loading factor of $\eta \sim 0.81$, by summing the $\Dot{M}_{\rm out}$ in all spaxels and dividing by total SFR. However, we strongly urge caution, that because our study is limited to optical ionised gas this number is a low-estimate. We speculate on the full mass-loading factor in the discussion. 

From spaxel-to-spaxel the value of $\eta$ varies by 2~orders-of-magnitude. Moreover, there is considerable systematic uncertainty introduced by the assumptions above. Varying the size of $R_{\rm out}$ changes the mass-loading factor to $\sim$0.2 and $\sim$1.2 at the larger and smaller $R_{\rm out}$ bounds respectively. 

We note that the inverse correlation in Fig.~\ref{fig:mlf_hbeta_sig_sfr} may be, at least in part, due to a circularity in that $\eta \propto 1/F_{\rm H\beta}$, which is also used to determine \sigmasfr. 
We therefore investigate what property drives the change in $\eta$. There are three measured quantities that we use to calculate $\eta$.  The H$\beta$ broad-to-narrow flux ratio $F_{\rm broad}/F_{\rm narrow}$, and \vout\ from the [OIII]~$\lambda5007$ line, which can be further broken into the velocity difference $v_{\rm diff}=|v_{\rm narrow}-v_{\rm broad}|$ and the outflow velocity dispersion $\sigma_{\rm broad}$. For velocity we find lower correlation coefficients with $\eta$ of 0.10, $-0.33$ and 0.29 for \vout, $v_{\rm diff}$ and $\sigma_{\rm broad}$ respectively. 
We have plotted the relationship between H$\beta$ $F_{\rm broad}/F_{\rm narrow}$ and \sigmasfr\ in the top row of Fig.~\ref{fig:mlf_hbeta_sig_sfr} for comparison to $\eta$. We see similar behaviour in both $\eta$ and H$\beta$ $F_{\rm broad}/F_{\rm narrow}$. We find that there is a much stronger correlation ($R\sim0.97$) between H$\beta$ $F_{\rm broad}/F_{\rm narrow}$ and $\eta$ than there is with either of the components of \vout. 
We conclude that the main driver of $\eta$ is the observed H$\beta$ $F_{\rm broad}/F_{\rm narrow}$.

In the lower panels of Figure \ref{fig:mlf_hbeta_sig_sfr} we plot the relationship $\eta~\propto$~\sigmasfr$^{-0.44}$ from simulation \citep{kim2020firstresultssmaug}, in which outflows are driven primarily by energy-driven winds similarly to the dotted line in Fig.~\ref{fig:sigma_sfr_vel_max}. \cite{li2017supernovaedriven}  similarly find $\eta~\propto$~\sigmasfr$^{-0.43}$, whereas \cite{creasey2013supernova} find a slightly steeper relationship of $\eta~\propto$~\sigmasfr$^{-0.74}$. 

In IRAS08, as shown in Fig.~\ref{fig:mlf_hbeta_sig_sfr}, 
we find a roughly constant $\eta$ across \sigmasfr\ 
in all three selection panels of the figure. The median of points in the left panel 
is $\eta\approx0.52$ with a root-mean-square scatter of 0.8. This increases to $\eta\approx0.7$ with a root-mean square scatter of 0.8 in the right panel. 
The overall trend we find in our data for \textit{strongly likely} fits (right panel) is consistent with the relationship found by \citet{kim2020firstresultssmaug}. 

Our overall trend in Fig.~\ref{fig:mlf_hbeta_sig_sfr} is different than that measured in galaxies at $z\approx 2$ \citep{newman2012sins,davies2019kiloparsec}. Both of those studies find a decrease in $\eta$ for low \sigmasfr. Moreover, we find no evidence of stark change in $\eta$ at \sigmasfr~$=1~$\sigmasfrunits, as suggested by \cite{newman2012sins}. This could be due to a unique aspect of our target, or alternatively may be due to uncertainties in measurements of the distant galaxies observed with adaptive optics. We note, also, that the SINS data does not reach the low \sigmasfr\ values, or the spatial resolution, of our nearby target.


Using NaD absorption line measurements from stacks of MaNGA data, \cite{robertsborsani2020outflowsMANGA} found a roughly constant relationship between $\eta$ and \sigmasfr\ for outflows located inside of twice the half-light radius, over a similar range in  \sigmasfr\ as ours. They also found an increase in $\eta$ for low \sigmasfr\ values at radii larger than 2$\times r_{1/2}$.  
It is difficult to compare the different tracers of the outflow mass-loading factor. NaD is not covered in our data. A systematic point-to-point comparison of outflows measured with different tracers of gas is direly needed.

\begin{figure*}
    \centering
    \includegraphics[width=\textwidth]{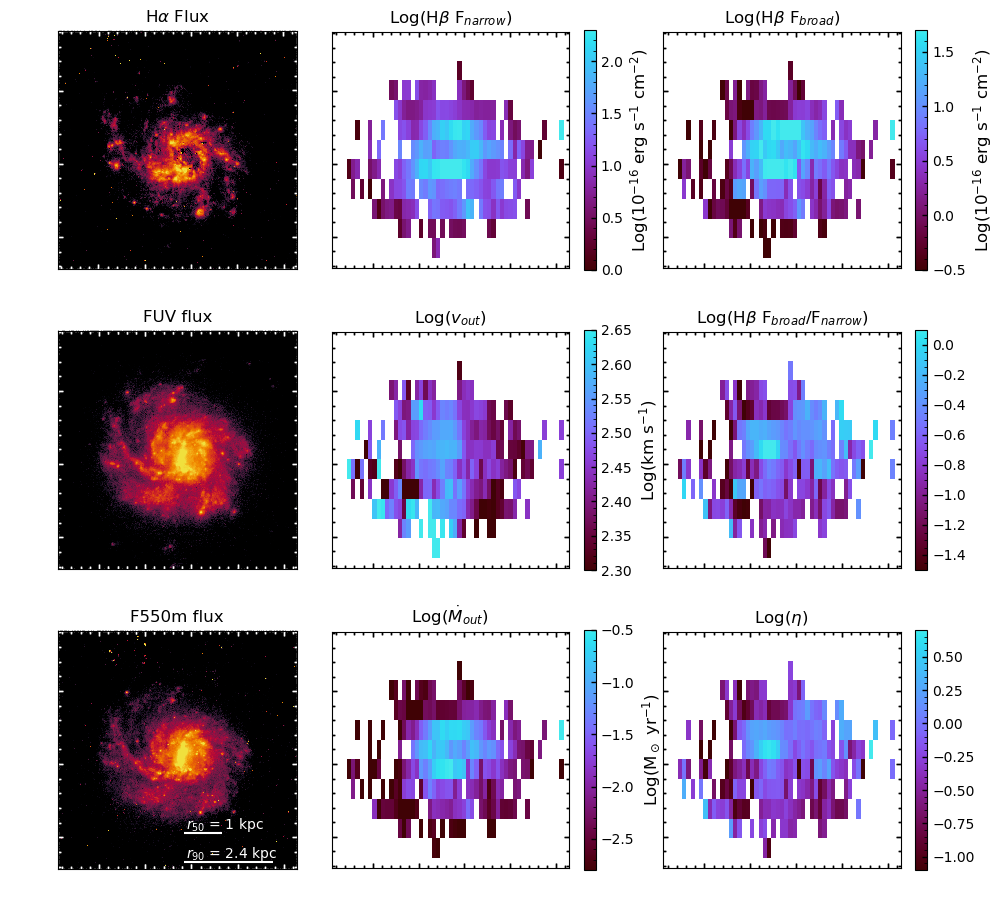}
    \caption{Here we map results from our observations of IRAS08: H$\beta$ narrow line flux from our fitting routine (\textit{top row middle}), H$\beta$ broad line flux from our fitting routine (\textit{top row right}), maximum outflow velocity \vout\ calculated using the [OIII]~$\lambda5007$ line (\textit{middle row middle}), H$\beta$ broad-to-narrow flux ratio F$_{\rm broad}$/F$_{\rm narrow}$ (\textit{middle row right}), mass outflow rate $\Dot{M}_{\rm out}$ (\textit{bottom row middle}), and mass loading factor $\eta$ (\textit{bottom row right}). Maps of H$\alpha$, FUV, and F550m flux from $HST$ are also given for comparison, with no extinction correction (\textit{left-hand panels of the top, middle and bottom rows respectively}). Note that the H$\beta$ emission line has a lower S/N than the [OIII] line, preventing an outflowing component from being resolved in some spaxels.  This means that quantities calculated using the H$\beta$ line have 9 fewer spaxels showing in the maps.}
    \label{fig:maps}
\end{figure*}

\subsection{Maps of outflow properties}
In Figure~\ref{fig:maps} we show maps of outflow properties from our KCWI observations and HST images of the ionised gas and starlight. The HST images have not been extinction corrected, however IRAS08 has low extinction ($<0.5$~mag). Our \vout\ is calculated using the fits to the [OIII]~$\lambda5007$ line (Eq.~\ref{eq:outflow_vel}).  Other than in the \vout\ panel (middle), all panels based on KCWI data use the H$\beta$ flux.  Due to the lower signal-to-noise of the H$\beta$ emission line in IRAS08, there are 9 fewer spaxels in which we are able to resolve the outflowing component for these maps.  
Note that $\Dot{M}_{\rm out}$ uses the velocity from [OIII] and luminosity from H$\beta$ (Eq.~\ref{eq:mass_outflow_rate}) and that \vout\ from H$\beta$ is consistent with that of [OIII] to $\sim$10\%, but [OIII] has higher S/N.

We find that in IRAS08, outflows are ubiquitous inside $r_{50}$ (1~kpc) with 100\% of spaxels containing outflows. This decreases at larger radius to 64\% of spaxels containing outflows from $r_{50}$ to $r_{90}$ (1~kpc to 2.4~kpc).  This trend is most clearly illustrated in the \vout\ panel of Fig.~\ref{fig:maps}.  Inside $r_{90}$ (2.4~kpc) 70\% of all spaxels contain outflows.  A recent survey of MaNGA galaxies with a broad range of SFRs found only 7\% contain detectable ionised gas outflows, with only roughly 1 spaxel per galaxy with an outflow \citep{RodriguezdelPino2019ionisedoutflowsmanga}. It is important to note that the MaNGA data has significantly reduced sensitivity compared to our KCWI data. Nonetheless, our results are consistent with outflows being widespread in disks with high \sigmasfr. 
 
For IRAS08 we calculate a total $\Dot{M}_{\rm out}\approx 7.9$~M$_\odot$~yr$^{-1}$. This is an order of magnitude greater than the $\Dot{M}_{\rm out}$ calculated by \cite{chisholm2017mass} of $\sim$0.33~M$_\odot$~yr$^{-1}$ using UV absorption lines with HST/COS.  We note that we sum up a larger area of the galaxy than is included in the 2.5\arcsec\ diameter ($\sim1$~kpc) COS pointing used by \cite{chisholm2017mass}.  If we sum up an area of similar size to a COS pointing, we find $\Dot{M}_{\rm out}\approx 6.0$~M$_\odot$~yr$^{-1}$.  All spaxels included in our total $\Dot{M}_{\rm out}$ are shown in the bottom middle panel of Fig.~\ref{fig:maps} and cover an area with a diameter $\sim4$~kpc.  These spaxels cover 70\% of the total galaxy surface area.


The vast majority of the outflowing mass originates from the star forming ring, not from the galaxy centre.  If we combine the spaxels containing the galaxy ring, we measure a $\Dot{M}_{\rm out}$ that is 60\% of the total outflowing mass from the entire galaxy located within 8\% of the total galaxy surface area. 
This has significant implications for the interpretation of single slit observations of outflows. Placing a slit simply on the galaxy centre may not necessarily capture the bulk of the outflowing mass. Parameter correlations with outflow properties for the entire galaxy \citep[e.g.][]{heckman2015systematic,chisholm2015scaling} may therefore have increased scatter from this effect. 

Because the calculation of $\Dot{M}_{\rm out}$ depends on multiple observables, we check that this peak in $\eta$ around the galaxy ring is reflected in the observed emission lines. In the top right panel of Fig.~\ref{fig:maps}, 
we show that total H$\beta$ flux from the broad spectral component is co-located within the region of the strong $\Dot{M}_{\rm out}$. We find a similar result with the [OIII] line. 

\section{Summary and Discussion} \label{sec:summary}
We have made resolved measurements of IRAS08 multi-component emission lines using observations from KCWI/Keck. We interpret the broad component as indicative of an outflow.  Using a fitting method that incorporates statistical and physically motivated tests, we have measured the kinematics and distribution of outflows across the disk in IRAS08.  We have two main results:  firstly, we find that our trends of \sigmasfr~vs.~\vout\ and \sigmasfr~vs.~$\eta$ are both broadly consistent with models of energy-driven winds. Moreover, we find this correlation becomes more robust at spatial scales between 0.6-1~kpc. Secondly, outflows are not limited to the galaxy centre, but are found with varying frequency at all radii within the 90\% radius of star light.  Indeed, the bulk of outflowing mass  is not colocated with the peak of galaxy emission. We discuss the implications of these results in the following subsections.

\subsection{Fitting Method: \textsc{koffee}}
We develop a method for fitting outflows in 
galaxies resolved to sub-kpc level. The problem of measuring resolved outflows presents the challenge of both deciding in which spaxels to use the extra component, and determining proper initial choices that facilitate the true $\chi^2$ minimum in traditional regression packages.  We apply a series of tests to determine whether fitting two Gaussians is justified in each spaxel. 
Making the simple assumption that all spaxels contain outflows leads to biases in measurements of outflows. The most significant is the presence of a large number of low velocity outflows, and the significant increase in the outflows that have very high broad-to-narrow flux ratios. Not incorporating similar tests for both outflow parameter estimation and the necessity of multiple Gaussian components would lead to a over estimation of physically interpreted quantities like the mass-loading factor.


\subsection{Comparison to Theory: Outflow Velocity}

It is widely accepted that the relationship between \vout\ and \sigmasfr\ provides constraints on the driving mechanisms behind the outflows.  Two popular models are energy-driven outflows, where \vout~$\propto$~\sigmasfr$^{0.1}$ \citep[e.g.][]{chen2010absorption,li2017supernovaedriven,kim2020firstresultssmaug, li2020simpleyetpowerful} and momentum-driven outflows, where \vout~$\propto$~\sigmasfr$^{2}$ \citetext{e.g. \citealp{murray2011radiationpressure}; see also discussion in \citealp{kornei2012properties}}.   Our results from IRAS08 are consistent with shallow powerlaws, as shown in Fig.~\ref{fig:sigma_sfr_vel_max} and Fig.~\ref{fig:sig_sfr_out_vel_binned}. 
There is currently not a clear consensus in the literature on the exact slope of the \vout$-$\sigmasfr\ relationship when measured for global galaxies. A number of studies do show shallow slopes \citep[e.g.][]{heckman2015systematic, RodriguezdelPino2019ionisedoutflowsmanga, robertsborsani2020outflowsMANGA} alternatively others do not find a robust correlation \citep[e.g.][]{Rubin2014evidence,chen2010absorption}. The important distinction between our observations and galaxy integrated studies is that our method removes an implicit correlation with galaxy mass, which is strongly correlated with observed outflow properties \citep[e.g.][]{chisholm2015scaling, heckman2015systematic}. Resolved studies, such as ours, are therefore an important step forward in fully characterising the dependency of outflow kinematics on properties of the underlying stellar population.

Using stacked $\sim$1~kpc spaxels from the SINS sample, \citet{davies2019kiloparsec} found a steeper relationship than ours, such that \vout~$\propto$~\sigmasfr$^{0.3}$ for ionised gas. They argued that because this result lies between the energy- and momentum-driven models, outflows in their data are driven by a combination of energy and momentum mechanisms. However, comparing our resolved results directly to those from stacked galaxy studies \citep[e.g.][]{davies2019kiloparsec, robertsborsani2020outflowsMANGA} may be too simplistic. Stacking galaxies may have systematic biases in comparison to individual resolved targets. We do not find that degrading the resolution to $\sim$1~kpc scales results in a different correlation. Indeed, the power-law remains shallow for all sampling scales we probe.  Moreover, very different flux sensitivities could lead to differences in returned outflow properties, for example high velocity low mass winds. It is not simple to understand how these would propagate through stacks. We note that if we average our results to make a single galaxy data point, then it does fall within the range of \citet{davies2019kiloparsec}'s 1-2~kpc-scaled data. Future work with JWST will likely prove informative to resolved outflows at $z>1$.

Overall we find that the \vout$-$\sigmasfr\ relationship in IRAS08 is consistent with those theories and simulations that drive outflows via energy from supernovae, rather than those that drive outflows via momentum from radiation by young stars.

\subsection{Comparison to Theory: Mass Loading Factor}

Similar to the velocity, the relationship between $\eta$ and \sigmasfr\ can place constraints on the driving mechanisms behind the outflows. The mass loading factor, $\eta$, is a measure of how efficiently the outflowing gas is coupled to the energy produced through the star formation process. Values of $\eta~\sim~1$ indicate that the mass of outflowing gas is comparable to the mass of gas being converted into stars.  
Low values of $\eta$ indicate that much more gas is converted into stars than is expelled in outflows, meaning that the energy from star formation is inefficiently coupled to the outflowing gas, or the star formation activity is insufficient to drive an outflow.  

We note that the total mass of outflows is multiphase in nature \citep[e.g.][]{fluetsch2020propertiesmultiphase, herrera-camus2020AGNfeedback}, and work on local starbursts and spirals suggests that the molecules dominate the outflow mass budget. The total value of $\eta$ in this work, therefore, significantly underestimates the total mass of the outflow. For example, \cite{bolatto2013suppression} have found $\eta\sim1-10$ in molecular phase gas in local starbursts with similar \sigmasfr\ as our target. In a galaxy at $z\sim5$ with similar \sigmasfr\ to IRAS08, \cite{herrera-camus2021kiloparsecview} found very high values of $\eta$ in spatially resolved [CII]. \cite{fluetsch2020propertiesmultiphase} compared ionised gas outflows to those of molecules in 4 galaxies with high SFR/M$_{\rm star}$, similar to IRAS08. They found roughly equal $\dot{M}_{\rm out}$ in the phases. This would imply $\eta\sim2$ in IRAS08. Alternatively \cite{robertsborsani2020observational} found that ions contribute a much lower fraction by mass, of order $\sim$1\% in nearby spirals of significantly lower SFR. This would make the mass-loading factor in IRAS08 very high. We therefore assume a reasonable total $\eta$ for IRAS08 is of order $\eta\sim2-10$.  Work directly comparing resolved outflows in both ions and molecules of starbursts is needed to interpret observations such as ours and those of high-z disks \citep[e.g.][]{davies2019kiloparsec}.

\begin{figure}
    \centering
    \includegraphics[width=\columnwidth]{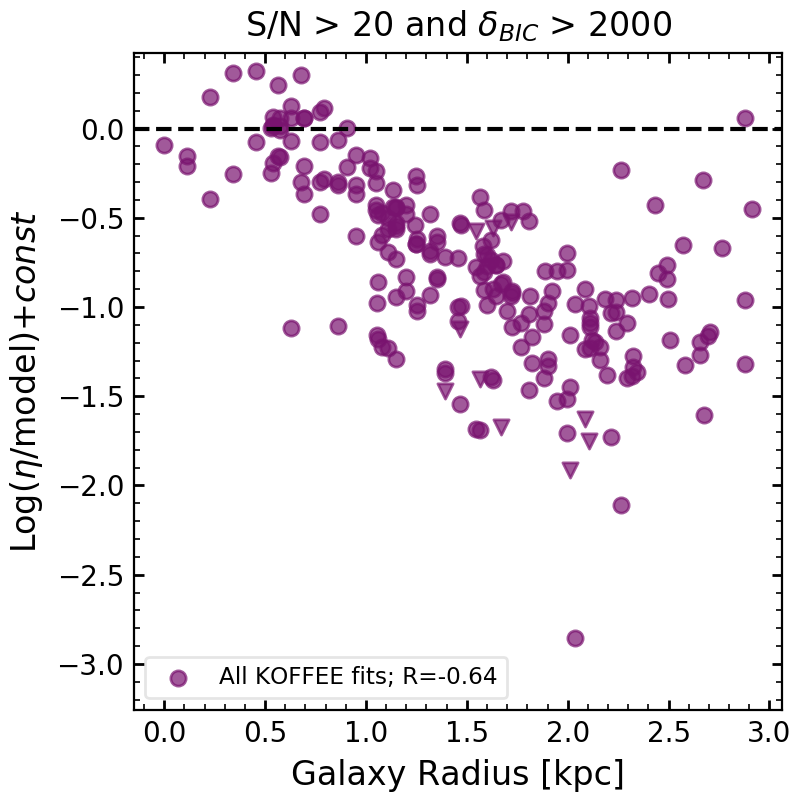}
    \caption{The y-axis shows the mass loading factor $\eta$ divided by the expected values for $\eta$ for their measured \sigmasfr\ from the supernovae driven wind model described in \citet{kim2020firstresultssmaug}, offset to zero at the galaxy centre by a constant. This is the same model which is shown in Fig.~\ref{fig:mlf_hbeta_sig_sfr}. The x-axis is the galaxy radius of each spaxel in kiloparsecs.  
    The Pearson's correlation coefficient, R, is given in the legend. 
    Points above (below) 0.0 indicate the model underestimated (overestimated) the $\eta$ value.  We find a strong trend of increasing overestimation of $\eta$ by the model compared to the observed data with increasing galaxy radius.}
    \label{fig:mlf_model_radius}
\end{figure}

Our observations of IRAS08 find a slight inverse correlation of $\eta$ with \sigmasfr\ for the \textit{highly likely} outflow fits, and no correlation for the remainder of the data (Fig.~\ref{fig:mlf_hbeta_sig_sfr}). 
Comparing to simulations, our results 
are consistent with those found by \citet{li2017supernovaedriven} and \citet{kim2020firstresultssmaug}.  
Other models incorporating energy-driven winds have also given negative slopes, under typical assumptions \citep[e.g.][]{creasey2013supernova}. 

Our slight inverse correlation of $\eta-$\sigmasfr\ is different from most observational studies, though it is difficult to say how these differences may depend on the method. Unlike with velocity, calculating $\eta$ requires normalising by the SFR. This may introduce systematic differences between resolved studies, stacks and global galaxy measurements.  A positive correlation between $\eta$ and \sigmasfr\ was found for stacks of SINS galaxies \citep{newman2012sins, davies2019kiloparsec} and for SDSS galaxies \citep{chen2010absorption}.  

In Figure \ref{fig:mlf_model_radius} we make a more detailed comparison between the $\eta$ in IRAS08 and that predicted by the models from \citet{kim2020firstresultssmaug}. 
Note that we have adjusted the values to set $\log(\eta_{data}/\eta_{model})=0$ at the centre of the galaxy, and then consider the relative change.  We find that our values of $\eta$ are increasingly lower, with respect to the model, with increasing radius.  Using radial stacks of MaNGA data, \citet{robertsborsani2020outflowsMANGA} found that for fixed \sigmasfr, $\eta$ increased with increasing galaxy radius, in a way that is similar to IRAS08 (see also our Fig.~\ref{fig:maps}).  Their results suggest that a property other than star formation processes may be varying $\eta$ across the galaxy disk.

From Eq. \ref{eq:mass_outflow_rate}, the calculation of $\eta$ depends on \vout, the flux ratio $F_{\rm broad}/F_{\rm narrow}$, and inversely depends on the outflow radius $R_{\rm out}$ and electron density $n_e$. Our results for \vout\ agree well with the model for energy-driven outflows (Fig. \ref{fig:sigma_sfr_vel_max}), and thus \vout\ is not likely driving the behaviour in Fig.~\ref{fig:mlf_model_radius}. We note that  \cite{kim2020firstresultssmaug} used kpc-scale box simulations and may not take into account possible differences in sub-galactic environments.  In the following paragraphs we consider possible assumptions for each of the remaining parameters that could lead to our result. 

Firstly, a very simple reason could be the available gas in the surrounding region of the galaxy. At the edge of a galaxy the local gas mass surface density is lower. A supernova explosion at the outskirts of the disk might have less gas available to entrain as it moves through the disk. This could then decrease the observed $F_{\rm broad}/F_{\rm narrow}$ without significant changes to the underlying physics. 



$R_{\rm out}$ is among the least well constrained quantities in estimating $\eta$. In this work, we assume a constant $R_{\rm out}$ across all spaxels.  The true value of $R_{\rm out}$ may, in fact, change from spaxel-to-spaxel across the face of IRAS08.  Nonetheless, we could make the {\em ad hoc} assumption that $R_{\rm out}$ increases with increasing galactocentric radius. This could, for example, occur if the $R_{\rm out}$ were coupled to local disk thickness, which increases with radius \citep[e.g.][]{garciadelacruz2021flaringthickdiscs}. Alternatively, one could make the {\em ad hoc} assumption that large \sigmasfr\ generates larger $R_{\rm out}$. It is, however, difficult to understand how $R_{\rm out}$ would vary with \sigmasfr\ without also changing \vout.

Another possibility is that the assumption of a constant $n_e$ across the face of the galaxy is incorrect. It seems plausible that more energetic outflows could alter the physical properties of the entrained gas, such as temperature and $n_e$. In principle this could be checked with follow up observations with higher spectral resolution in the [OII] doublet or alternatively a separate instrument capable of measuring the [SII] doublet, which is outside of KCWI's wavelength range. As is discussed in previous  works \citep[e.g.][]{davies2019kiloparsec}, the $n_e$ of outflows is poorly constrained, especially in these strong wind systems. 

\vskip 5pt

Altogether, our results in IRAS08 for both \vout$-$\sigmasfr\ and $\eta-$\sigmasfr\ relations are consistent with models for energy-driven outflows.  
We note that we have only discussed consistency with the powerlaw slope. For example, the simulations of \cite{kim2020firstresultssmaug} have found a similar powerlaw in the outflow velocity, but only produced outflows with velocity of order $\sim$100~km~s$^{-1}$ at the same \sigmasfr\ as our target. Historically, it has been noted that supernova-driven outflows, under standard assumptions, do not produce as high velocity winds as has been observed in starbursting galaxies \citep{fielding2018clustered}. 
The precise definition of \vout\ used may introduce systematics at the factor of a few level. An apples-to-apples comparison to mock observations would be useful in this case.



\subsection{Implications of the Spatial distribution and Covering fraction of outflows}
Determining the mass-outflow rate for an entire galaxy requires estimating the fraction of the star light that is covered by outflows. In large surveys of unresolved galaxies this is almost always an assumed quantity, often with guidance from the residual flux of saturated absorption lines \citep[review][]{veilleux2020cooloutflows}. Our observations can give guidance on covering fractions of star bursting disks. 

Outflows in disk galaxies in our local universe are typically thought to have biconical outflows \citep[e.g.][]{bland1988m82, shopbell1998asymmetricwindM82}, where the base of the outflow is narrowly focused on the centre of the galaxy.  However, from our observations of IRAS08 we have found the signature of outflows across the entire galaxy disk. Within $r_{50}$ we find $\sim$100\% of the disk has detectable outflows, and $\sim70$\% of the area within $r_{90}$. This is consistent with absorption line measurements from \citet{chisholm2017mass} who find $\sim80\%$ on gas within $r_{50}$ of IRAS08. Our results suggest that disk galaxies with high \sigmasfr\ will not have the same distribution of winds as nearby starbursts. If one defines the covering fraction simply as the ratio of the continuum which has associated outflows, our results imply large covering fractions, $C_f\approx 0.7-1$, for disk galaxies with high SFR surface density, as is common above $z>0.5$.   

In IRAS08, $\Dot{M}_{\rm out}$ measurements are heavily concentrated in a small region of the galaxy, corresponding to the star forming ring contained with a radius of $\sim$0.5~kpc of the galaxy centre (Fig.~\ref{fig:maps}). 
Indeed, 60\% of the total outflow mass is colocated with this ring, which only covers 8\% of the total area of the star-light within $r_{90}$. If one observed IRAS08 with low spatial resolution the outflow mass might reflect this smaller region. The flux measured from only the ring would therefore make a small difference on the mass outflow rate, but the covering fraction might change by an order-of-magnitude.


There has been debate about whether outflows in starbursting galaxies at high-z are galaxy-wide outflows or whether they are restricted to the high surface brightness clumps of star formation. For example, \cite{bordoloi2016spatiallyresolved} placed slits on clumps of a lensed target, and found outflows preferentially located there. Similarly, \cite{newman2012sins} found outflows restricted to high surface brightness clumps. Our results in IRAS08 imply a balanced take. Outflows are found to be wide spread in the disk, but large fractions of that gas may come from small regions. We find no evidence in IRAS08 for a cut-off or reduction in outflow frequency at \sigmasfr~$\sim0.1$~M$_{\odot}$~yr$^{-1}$~kpc$^{-2}$. One simple way to reconcile our observation with those at $z>1$ would be that the fainter outflows are more challenging to observe with current telescopes. Observations from the upcoming JWST and further in the future ELT are likely to determine if this is the case.

\vskip 10pt
In summary we found that the scaling relationships of outflow properties with the colocated \sigmasfr\ is in general consistent with predictions in which outflows are primarily driven by supernovae. However, this is only one target and differences may occur galaxy to galaxy, which must be accounted for. Future work applying our technique to more targets is needed to test these results. We will apply our technique to the face-on targets in the DUVET survey in a future paper. 

\section*{Acknowledgements}

The authors would like to thank the referee for the time and effort they spent to give useful comments which improved the paper. 
Parts of this research were supported by the Australian Research Council Centre of Excellence for All Sky Astrophysics in 3 Dimensions (ASTRO 3D), through project number CE170100013.
D.B.F. acknowledges support from Australian Research Council (ARC)  Future Fellowship FT170100376. N.M.N.~and G.G.K.~acknowledge the support of the Australian Research Council through Discovery Project grant DP170103470. 
R.H.-C. thanks the Max Planck Society for support under the Partner Group project "The Baryon Cycle in Galaxies" between the Max Planck for Extraterrestrial Physics and the Universidad de Concepción. R.H-C. also gratefully acknowledge financial support from Millenium Nucleus NCN19058 (TITANs), and ANID BASAL projects ACE210002 and FB210003.
K.S. and R.R.V. acknowledge funding support from National Science Foundation Award No. 1816462.
M.G. is grateful to the Fonds de recherche du Qu\'ebec-Nature et Technologies (FRQNT) for financial support.
The data presented herein were obtained at the W.~M.~Keck Observatory, which is operated as a scientific partnership among the California Institute of Technology, the University of California and the National Aeronautics and Space Administration. The Observatory was made possible by the generous financial support of the W. M. Keck Foundation. Observations were supported by Swinburne Keck program 2018A\_W185. The authors wish to recognise and acknowledge the very significant cultural role and reverence that the summit of Maunakea has always had within the indigenous Hawaiian community. We are most fortunate to have the opportunity to conduct observations from this mountain.

\section*{Data Availability}

The DUVET Survey is still in progress.  The data underlying this article will be shared on reasonable request to the PI, Deanne Fisher at dfisher@swin.edu.au



\bibliographystyle{mnras}
\bibliography{bibliography} %

\begin{thebibliography}{}
\makeatletter
\relax
\def\mn@urlcharsother{\let\do\@makeother \do\$\do\&\do\#\do\^\do\_\do\%\do\~}
\def\mn@doi{\begingroup\mn@urlcharsother \@ifnextchar [ {\mn@doi@}
  {\mn@doi@[]}}
\def\mn@doi@[#1]#2{\def\@tempa{#1}\ifx\@tempa\@empty \href
  {http://dx.doi.org/#2} {doi:#2}\else \href {http://dx.doi.org/#2} {#1}\fi
  \endgroup}
\def\mn@eprint#1#2{\mn@eprint@#1:#2::\@nil}
\def\mn@eprint@arXiv#1{\href {http://arxiv.org/abs/#1} {{\tt arXiv:#1}}}
\def\mn@eprint@dblp#1{\href {http://dblp.uni-trier.de/rec/bibtex/#1.xml}
  {dblp:#1}}
\def\mn@eprint@#1:#2:#3:#4\@nil{\def\@tempa {#1}\def\@tempb {#2}\def\@tempc
  {#3}\ifx \@tempc \@empty \let \@tempc \@tempb \let \@tempb \@tempa \fi \ifx
  \@tempb \@empty \def\@tempb {arXiv}\fi \@ifundefined
  {mn@eprint@\@tempb}{\@tempb:\@tempc}{\expandafter \expandafter \csname
  mn@eprint@\@tempb\endcsname \expandafter{\@tempc}}}

\bibitem[\protect\citeauthoryear{{Arribas}, {Colina}, {Bellocchi}, {Maiolino}
  \& {Villar-Mart{\'\i}n}}{{Arribas}
  et~al.}{2014}]{arribas2014ionisedgasoutflows}
{Arribas} S.,  {Colina} L.,  {Bellocchi} E.,  {Maiolino} R.,
  {Villar-Mart{\'\i}n} M.,  2014, \mn@doi [\aap] {10.1051/0004-6361/201323324},
  \href {https://ui.adsabs.harvard.edu/abs/2014A&A...568A..14A} {568, A14}

\bibitem[\protect\citeauthoryear{{Avery} et~al.,}{{Avery}
  et~al.}{2021}]{avery2021incidence}
{Avery} C.~R.,  et~al., 2021, \mn@doi [\mnras] {10.1093/mnras/stab780}, \href
  {https://ui.adsabs.harvard.edu/abs/2021MNRAS.tmp..787A} {}

\bibitem[\protect\citeauthoryear{{Bland} \& {Tully}}{{Bland} \&
  {Tully}}{1988}]{bland1988m82}
{Bland} J.,  {Tully} B.,  1988, \mn@doi [\nat] {10.1038/334043a0}, \href
  {https://ui.adsabs.harvard.edu/abs/1988Natur.334...43B} {334, 43}

\bibitem[\protect\citeauthoryear{{Bolatto} et~al.,}{{Bolatto}
  et~al.}{2013}]{bolatto2013suppression}
{Bolatto} A.~D.,  et~al., 2013, \mn@doi [\nat] {10.1038/nature12351}, \href
  {https://ui.adsabs.harvard.edu/abs/2013Natur.499..450B} {499, 450}

\bibitem[\protect\citeauthoryear{{Bordoloi}, {Rigby}, {Tumlinson}, {Bayliss},
  {Sharon}, {Gladders}  \& {Wuyts}}{{Bordoloi}
  et~al.}{2016}]{bordoloi2016spatiallyresolved}
{Bordoloi} R.,  {Rigby} J.~R.,  {Tumlinson} J.,  {Bayliss} M.~B.,  {Sharon} K.,
   {Gladders} M.~G.,   {Wuyts} E.,  2016, \mn@doi [\mnras]
  {10.1093/mnras/stw449}, \href
  {https://ui.adsabs.harvard.edu/abs/2016MNRAS.458.1891B} {458, 1891}

\bibitem[\protect\citeauthoryear{{Calzetti}}{{Calzetti}}{2001}]{calzetti2001dustopacity}
{Calzetti} D.,  2001, \mn@doi [\pasp] {10.1086/324269}, \href
  {https://ui.adsabs.harvard.edu/abs/2001PASP..113.1449C} {113, 1449}

\bibitem[\protect\citeauthoryear{{Calzetti}, {Armus}, {Bohlin}, {Kinney},
  {Koornneef}  \& {Storchi-Bergmann}}{{Calzetti}
  et~al.}{2000}]{calzetti2000dustcontent}
{Calzetti} D.,  {Armus} L.,  {Bohlin} R.~C.,  {Kinney} A.~L.,  {Koornneef} J.,
   {Storchi-Bergmann} T.,  2000, \mn@doi [\apj] {10.1086/308692}, \href
  {https://ui.adsabs.harvard.edu/abs/2000ApJ...533..682C} {533, 682}

\bibitem[\protect\citeauthoryear{{Cannon}, {Skillman}, {Kunth}, {Leitherer},
  {Mas-Hesse}, {{\"O}stlin}  \& {Petrosian}}{{Cannon}
  et~al.}{2004}]{cannon2004extendedtidalstructure}
{Cannon} J.~M.,  {Skillman} E.~D.,  {Kunth} D.,  {Leitherer} C.,  {Mas-Hesse}
  M.,  {{\"O}stlin} G.,   {Petrosian} A.,  2004, \mn@doi [\apj]
  {10.1086/420868}, \href
  {https://ui.adsabs.harvard.edu/abs/2004ApJ...608..768C} {608, 768}

\bibitem[\protect\citeauthoryear{{Cappellari}}{{Cappellari}}{2017}]{cappellari2017ppxf}
{Cappellari} M.,  2017, \mn@doi [\mnras] {10.1093/mnras/stw3020}, \href
  {https://ui.adsabs.harvard.edu/abs/2017MNRAS.466..798C} {466, 798}

\bibitem[\protect\citeauthoryear{{Cardelli}, {Clayton}  \& {Mathis}}{{Cardelli}
  et~al.}{1989}]{cardelli1989relationship}
{Cardelli} J.~A.,  {Clayton} G.~C.,   {Mathis} J.~S.,  1989, \mn@doi [\apj]
  {10.1086/167900}, \href
  {https://ui.adsabs.harvard.edu/abs/1989ApJ...345..245C} {345, 245}

\bibitem[\protect\citeauthoryear{Chen, Tremonti, Heckman, Kauffmann, Weiner,
  Brinchmann  \& Wang}{Chen et~al.}{2010}]{chen2010absorption}
Chen Y.-M.,  Tremonti C.~A.,  Heckman T.~M.,  Kauffmann G.,  Weiner B.~J.,
  Brinchmann J.,   Wang J.,  2010, The Astronomical Journal, 140, 445

\bibitem[\protect\citeauthoryear{Chisholm, Tremonti, Leitherer, Chen, Wofford
  \& Lundgren}{Chisholm et~al.}{2015}]{chisholm2015scaling}
Chisholm J.,  Tremonti C.~A.,  Leitherer C.,  Chen Y.,  Wofford A.,   Lundgren
  B.,  2015, The Astrophysical Journal, 811, 149

\bibitem[\protect\citeauthoryear{{Chisholm}, {Tremonti Christy}, {Leitherer}
  \& {Chen}}{{Chisholm} et~al.}{2016}]{chisholm2016robust}
{Chisholm} J.,  {Tremonti Christy} A.,  {Leitherer} C.,   {Chen} Y.,  2016,
  \mn@doi [\mnras] {10.1093/mnras/stw1951}, \href
  {https://ui.adsabs.harvard.edu/abs/2016MNRAS.463..541C} {463, 541}

\bibitem[\protect\citeauthoryear{Chisholm, Tremonti, Leitherer  \&
  Chen}{Chisholm et~al.}{2017}]{chisholm2017mass}
Chisholm J.,  Tremonti C.~A.,  Leitherer C.,   Chen Y.,  2017, Monthly Notices
  of the Royal Astronomical Society, 469, 4831

\bibitem[\protect\citeauthoryear{{Creasey}, {Theuns}  \& {Bower}}{{Creasey}
  et~al.}{2013}]{creasey2013supernova}
{Creasey} P.,  {Theuns} T.,   {Bower} R.~G.,  2013, \mn@doi [\mnras]
  {10.1093/mnras/sts439}, \href
  {https://ui.adsabs.harvard.edu/abs/2013MNRAS.429.1922C} {429, 1922}

\bibitem[\protect\citeauthoryear{Davies et~al.,}{Davies
  et~al.}{2019}]{davies2019kiloparsec}
Davies R.~L.,  et~al., 2019, The Astrophysical Journal, 873, 122

\bibitem[\protect\citeauthoryear{{Ferrara} \& {Ricotti}}{{Ferrara} \&
  {Ricotti}}{2006}]{ferrara2006winds}
{Ferrara} A.,  {Ricotti} M.,  2006, \mn@doi [\mnras]
  {10.1111/j.1365-2966.2006.10978.x}, \href
  {https://ui.adsabs.harvard.edu/abs/2006MNRAS.373..571F} {373, 571}

\bibitem[\protect\citeauthoryear{Fielding, Quataert  \& Martizzi}{Fielding
  et~al.}{2018}]{fielding2018clustered}
Fielding D.,  Quataert E.,   Martizzi D.,  2018, Monthly Notices of the Royal
  Astronomical Society, 481, 3325

\bibitem[\protect\citeauthoryear{{Fisher} et~al.,}{{Fisher}
  et~al.}{2017}]{fisher2017DYNAMO-HSTsurvey}
{Fisher} D.~B.,  et~al., 2017, \mn@doi [\mnras] {10.1093/mnras/stw2281}, \href
  {https://ui.adsabs.harvard.edu/abs/2017MNRAS.464..491F} {464, 491}

\bibitem[\protect\citeauthoryear{Fisher, Bolatto, White, Glazebrook, Abraham
  \& Obreschkow}{Fisher et~al.}{2019}]{fisher2019testing}
Fisher D.~B.,  Bolatto A.~D.,  White H.,  Glazebrook K.,  Abraham R.~G.,
  Obreschkow D.,  2019, The Astrophysical Journal, 870, 46

\bibitem[\protect\citeauthoryear{{Fluetsch} et~al.,}{{Fluetsch}
  et~al.}{2020}]{fluetsch2020propertiesmultiphase}
{Fluetsch} A.,  et~al., 2020, arXiv e-prints, \href
  {https://ui.adsabs.harvard.edu/abs/2020arXiv200613232F} {p. arXiv:2006.13232}

\bibitem[\protect\citeauthoryear{{F{\"o}rster Schreiber} et~al.,}{{F{\"o}rster
  Schreiber} et~al.}{2011}]{forsterschreiber2011sinssurvey}
{F{\"o}rster Schreiber} N.~M.,  et~al., 2011, The Messenger, \href
  {https://ui.adsabs.harvard.edu/abs/2011Msngr.145...39F} {145, 39}

\bibitem[\protect\citeauthoryear{F{\"o}rster~Schreiber
  et~al.,}{F{\"o}rster~Schreiber et~al.}{2019}]{forsterschreiber2019kmos3d}
F{\"o}rster~Schreiber N.,  et~al., 2019, The Astrophysical Journal, 875, 21

\bibitem[\protect\citeauthoryear{{Garc{\'\i}a de la Cruz}, {Martig}, {Minchev}
  \& {James}}{{Garc{\'\i}a de la Cruz}
  et~al.}{2021}]{garciadelacruz2021flaringthickdiscs}
{Garc{\'\i}a de la Cruz} J.,  {Martig} M.,  {Minchev} I.,   {James} P.,  2021,
  \mn@doi [\mnras] {10.1093/mnras/staa3906}, \href
  {https://ui.adsabs.harvard.edu/abs/2021MNRAS.501.5105G} {501, 5105}

\bibitem[\protect\citeauthoryear{Genzel et~al.,}{Genzel
  et~al.}{2011}]{genzel2011sins}
Genzel R.,  et~al., 2011, The Astrophysical Journal, 733, 101

\bibitem[\protect\citeauthoryear{{Girard} et~al.,}{{Girard}
  et~al.}{2021}]{girard2021systematic}
{Girard} M.,  et~al., 2021, \mn@doi [\apj] {10.3847/1538-4357/abd5b9}, \href
  {https://ui.adsabs.harvard.edu/abs/2021ApJ...909...12G} {909, 12}

\bibitem[\protect\citeauthoryear{{Gonz{\'a}lez Delgado}, {Leitherer},
  {Heckman}, {Lowenthal}, {Ferguson}  \& {Robert}}{{Gonz{\'a}lez Delgado}
  et~al.}{1998}]{gonzalezdelgado1998FUVspectra}
{Gonz{\'a}lez Delgado} R.~M.,  {Leitherer} C.,  {Heckman} T.,  {Lowenthal}
  J.~D.,  {Ferguson} H.~C.,   {Robert} C.,  1998, \mn@doi [\apj]
  {10.1086/305321}, \href
  {https://ui.adsabs.harvard.edu/abs/1998ApJ...495..698G} {495, 698}

\bibitem[\protect\citeauthoryear{{Hao}, {Kennicutt}, {Johnson}, {Calzetti},
  {Dale}  \& {Moustakas}}{{Hao} et~al.}{2011}]{hao2011dustcorrected}
{Hao} C.-N.,  {Kennicutt} R.~C.,  {Johnson} B.~D.,  {Calzetti} D.,  {Dale}
  D.~A.,   {Moustakas} J.,  2011, \mn@doi [\apj] {10.1088/0004-637X/741/2/124},
  \href {https://ui.adsabs.harvard.edu/abs/2011ApJ...741..124H} {741, 124}

\bibitem[\protect\citeauthoryear{{Heckman}}{{Heckman}}{2002}]{heckman2002galacticsuperwinds}
{Heckman} T.~M.,  2002, in {Mulchaey} J.~S.,  {Stocke} J.~T.,  eds,
  Astronomical Society of the Pacific Conference Series Vol. 254, Extragalactic
  Gas at Low Redshift. p.~292 (\mn@eprint {arXiv} {astro-ph/0107438})

\bibitem[\protect\citeauthoryear{{Heckman} \& {Borthakur}}{{Heckman} \&
  {Borthakur}}{2016}]{heckman2016implications}
{Heckman} T.~M.,  {Borthakur} S.,  2016, \mn@doi [\apj]
  {10.3847/0004-637X/822/1/9}, \href
  {https://ui.adsabs.harvard.edu/abs/2016ApJ...822....9H} {822, 9}

\bibitem[\protect\citeauthoryear{{Heckman}, {Lehnert}, {Strickland }  \&
  {Armus}}{{Heckman} et~al.}{2000}]{heckman2000absorptionline}
{Heckman} T.~M.,  {Lehnert} M.~D.,  {Strickland } D.~K.,   {Armus} L.,  2000,
  \mn@doi [\apjs] {10.1086/313421}, \href
  {https://ui.adsabs.harvard.edu/abs/2000ApJS..129..493H} {129, 493}

\bibitem[\protect\citeauthoryear{Heckman, Alexandroff, Borthakur, Overzier  \&
  Leitherer}{Heckman et~al.}{2015}]{heckman2015systematic}
Heckman T.~M.,  Alexandroff R.~M.,  Borthakur S.,  Overzier R.,   Leitherer C.,
   2015, The Astrophysical Journal, 809, 147

\bibitem[\protect\citeauthoryear{{Herrera-Camus} et~al.,}{{Herrera-Camus}
  et~al.}{2020}]{herrera-camus2020AGNfeedback}
{Herrera-Camus} R.,  et~al., 2020, \mn@doi [\aap]
  {10.1051/0004-6361/201936434}, \href
  {https://ui.adsabs.harvard.edu/abs/2020A&A...635A..47H} {635, A47}

\bibitem[\protect\citeauthoryear{{Herrera-Camus} et~al.,}{{Herrera-Camus}
  et~al.}{2021}]{herrera-camus2021kiloparsecview}
{Herrera-Camus} R.,  et~al., 2021, \mn@doi [\aap]
  {10.1051/0004-6361/202039704}, \href
  {https://ui.adsabs.harvard.edu/abs/2021A&A...649A..31H} {649, A31}

\bibitem[\protect\citeauthoryear{{Hinshaw} et~al.,}{{Hinshaw}
  et~al.}{2013}]{hinshaw2013wmap9}
{Hinshaw} G.,  et~al., 2013, \mn@doi [\apjs] {10.1088/0067-0049/208/2/19},
  \href {https://ui.adsabs.harvard.edu/abs/2013ApJS..208...19H} {208, 19}

\bibitem[\protect\citeauthoryear{{Ho} et~al.,}{{Ho} et~al.}{2014}]{ho2014sami}
{Ho} I.~T.,  et~al., 2014, \mn@doi [\mnras] {10.1093/mnras/stu1653}, \href
  {https://ui.adsabs.harvard.edu/abs/2014MNRAS.444.3894H} {444, 3894}

\bibitem[\protect\citeauthoryear{{Hopkins}, {Quataert}  \& {Murray}}{{Hopkins}
  et~al.}{2012}]{hopkins2012stellarfeedback}
{Hopkins} P.~F.,  {Quataert} E.,   {Murray} N.,  2012, \mn@doi [\mnras]
  {10.1111/j.1365-2966.2012.20593.x}, \href
  {https://ui.adsabs.harvard.edu/abs/2012MNRAS.421.3522H} {421, 3522}

\bibitem[\protect\citeauthoryear{{Hopkins}, {Kere{\v{s}}}, {O{\~n}orbe},
  {Faucher-Gigu{\`e}re}, {Quataert}, {Murray}  \& {Bullock}}{{Hopkins}
  et~al.}{2014}]{hopkins2014FIRE}
{Hopkins} P.~F.,  {Kere{\v{s}}} D.,  {O{\~n}orbe} J.,  {Faucher-Gigu{\`e}re}
  C.-A.,  {Quataert} E.,  {Murray} N.,   {Bullock} J.~S.,  2014, \mn@doi
  [\mnras] {10.1093/mnras/stu1738}, \href
  {https://ui.adsabs.harvard.edu/abs/2014MNRAS.445..581H} {445, 581}

\bibitem[\protect\citeauthoryear{{Hung} et~al.,}{{Hung}
  et~al.}{2019}]{hung2019whatdrives}
{Hung} C.-L.,  et~al., 2019, \mn@doi [\mnras] {10.1093/mnras/sty2970}, \href
  {https://ui.adsabs.harvard.edu/abs/2019MNRAS.482.5125H} {482, 5125}

\bibitem[\protect\citeauthoryear{{Jones}, {Stark}  \& {Ellis}}{{Jones}
  et~al.}{2018}]{jones2018dust}
{Jones} T.,  {Stark} D.~P.,   {Ellis} R.~S.,  2018, \mn@doi [\apj]
  {10.3847/1538-4357/aad37f}, \href
  {https://ui.adsabs.harvard.edu/abs/2018ApJ...863..191J} {863, 191}

\bibitem[\protect\citeauthoryear{Kass \& Raftery}{Kass \&
  Raftery}{1995}]{Kass1995BayesFactors}
Kass R.~E.,  Raftery A.~E.,  1995, Journal of the American Statistical
  Association, 90, 773

\bibitem[\protect\citeauthoryear{{Kim}, {Ostriker}  \& {Kim}}{{Kim}
  et~al.}{2013}]{kim2013threeDhydrosim}
{Kim} C.-G.,  {Ostriker} E.~C.,   {Kim} W.-T.,  2013, \mn@doi [\apj]
  {10.1088/0004-637X/776/1/1}, \href
  {https://ui.adsabs.harvard.edu/abs/2013ApJ...776....1K} {776, 1}

\bibitem[\protect\citeauthoryear{{Kim} et~al.,}{{Kim}
  et~al.}{2020}]{kim2020firstresultssmaug}
{Kim} C.-G.,  et~al., 2020, \mn@doi [\apj] {10.3847/1538-4357/aba962}, \href
  {https://ui.adsabs.harvard.edu/abs/2020ApJ...900...61K} {900, 61}

\bibitem[\protect\citeauthoryear{{Kornei}, {Shapley}, {Martin}, {Coil}, {Lotz},
  {Schiminovich}, {Bundy}  \& {Noeske}}{{Kornei}
  et~al.}{2012}]{kornei2012properties}
{Kornei} K.~A.,  {Shapley} A.~E.,  {Martin} C.~L.,  {Coil} A.~L.,  {Lotz}
  J.~M.,  {Schiminovich} D.,  {Bundy} K.,   {Noeske} K.~G.,  2012, \mn@doi
  [\apj] {10.1088/0004-637X/758/2/135}, \href
  {https://ui.adsabs.harvard.edu/abs/2012ApJ...758..135K} {758, 135}

\bibitem[\protect\citeauthoryear{{Krieger} et~al.,}{{Krieger}
  et~al.}{2019}]{krieger2019molecularoutflow}
{Krieger} N.,  et~al., 2019, \mn@doi [\apj] {10.3847/1538-4357/ab2d9c}, \href
  {https://ui.adsabs.harvard.edu/abs/2019ApJ...881...43K} {881, 43}

\bibitem[\protect\citeauthoryear{{Krumholz}, {Burkhart}, {Forbes}  \&
  {Crocker}}{{Krumholz} et~al.}{2018}]{krumholz2018unifiedmodel}
{Krumholz} M.~R.,  {Burkhart} B.,  {Forbes} J.~C.,   {Crocker} R.~M.,  2018,
  \mn@doi [\mnras] {10.1093/mnras/sty852}, \href
  {https://ui.adsabs.harvard.edu/abs/2018MNRAS.477.2716K} {477, 2716}

\bibitem[\protect\citeauthoryear{{Leitherer}, {Li}, {Calzetti}  \&
  {Heckman}}{{Leitherer} et~al.}{2002}]{Leitherer2002globalFUV}
{Leitherer} C.,  {Li} I.~H.,  {Calzetti} D.,   {Heckman} T.~M.,  2002, \mn@doi
  [\apjs] {10.1086/342486}, \href
  {https://ui.adsabs.harvard.edu/abs/2002ApJS..140..303L} {140, 303}

\bibitem[\protect\citeauthoryear{{Leroy} et~al.,}{{Leroy}
  et~al.}{2015}]{leroy2015multiphaseM82}
{Leroy} A.~K.,  et~al., 2015, \mn@doi [\apj] {10.1088/0004-637X/814/2/83},
  \href {https://ui.adsabs.harvard.edu/abs/2015ApJ...814...83L} {814, 83}

\bibitem[\protect\citeauthoryear{{Li} \& {Bryan}}{{Li} \&
  {Bryan}}{2020}]{li2020simpleyetpowerful}
{Li} M.,  {Bryan} G.~L.,  2020, \mn@doi [\apjl] {10.3847/2041-8213/ab7304},
  \href {https://ui.adsabs.harvard.edu/abs/2020ApJ...890L..30L} {890, L30}

\bibitem[\protect\citeauthoryear{{Li}, {Bryan}  \& {Ostriker}}{{Li}
  et~al.}{2017}]{li2017supernovaedriven}
{Li} M.,  {Bryan} G.~L.,   {Ostriker} J.~P.,  2017, \mn@doi [\apj]
  {10.3847/1538-4357/aa7263}, \href
  {https://ui.adsabs.harvard.edu/abs/2017ApJ...841..101L} {841, 101}

\bibitem[\protect\citeauthoryear{{L{\'o}pez-S{\'a}nchez}, {Esteban}  \&
  {Garc{\'\i}a-Rojas}}{{L{\'o}pez-S{\'a}nchez}
  et~al.}{2006}]{lopezsanchez2006IRAS08paper}
{L{\'o}pez-S{\'a}nchez} {\'A}.~R.,  {Esteban} C.,   {Garc{\'\i}a-Rojas} J.,
  2006, \mn@doi [\aap] {10.1051/0004-6361:20053119}, \href
  {https://ui.adsabs.harvard.edu/abs/2006A&A...449..997L} {449, 997}

\bibitem[\protect\citeauthoryear{{Madau} \& {Dickinson}}{{Madau} \&
  {Dickinson}}{2014}]{madau2014cosmicSFH}
{Madau} P.,  {Dickinson} M.,  2014, \mn@doi [\araa]
  {10.1146/annurev-astro-081811-125615}, \href
  {https://ui.adsabs.harvard.edu/abs/2014ARA&A..52..415M} {52, 415}

\bibitem[\protect\citeauthoryear{{Martin}}{{Martin}}{2005}]{martin2005mapping}
{Martin} C.~L.,  2005, \mn@doi [\apj] {10.1086/427277}, \href
  {https://ui.adsabs.harvard.edu/abs/2005ApJ...621..227M} {621, 227}

\bibitem[\protect\citeauthoryear{{Morrissey} et~al.,}{{Morrissey}
  et~al.}{2018}]{Morrissey2018kcwi}
{Morrissey} P.,  et~al., 2018, \mn@doi [\apj] {10.3847/1538-4357/aad597}, \href
  {https://ui.adsabs.harvard.edu/abs/2018ApJ...864...93M} {864, 93}

\bibitem[\protect\citeauthoryear{{Mosleh}, {Williams}  \& {Franx}}{{Mosleh}
  et~al.}{2013}]{mosleh2013robustness}
{Mosleh} M.,  {Williams} R.~J.,   {Franx} M.,  2013, \mn@doi [\apj]
  {10.1088/0004-637X/777/2/117}, \href
  {https://ui.adsabs.harvard.edu/abs/2013ApJ...777..117M} {777, 117}

\bibitem[\protect\citeauthoryear{{Murray}, {M{\'e}nard}  \&
  {Thompson}}{{Murray} et~al.}{2011}]{murray2011radiationpressure}
{Murray} N.,  {M{\'e}nard} B.,   {Thompson} T.~A.,  2011, \mn@doi [\apj]
  {10.1088/0004-637X/735/1/66}, \href
  {https://ui.adsabs.harvard.edu/abs/2011ApJ...735...66M} {735, 66}

\bibitem[\protect\citeauthoryear{{Nelson} et~al.,}{{Nelson}
  et~al.}{2019}]{nelson2019firstresultsTNG50}
{Nelson} D.,  et~al., 2019, \mn@doi [\mnras] {10.1093/mnras/stz2306}, \href
  {https://ui.adsabs.harvard.edu/abs/2019MNRAS.490.3234N} {490, 3234}

\bibitem[\protect\citeauthoryear{Newman et~al.,}{Newman
  et~al.}{2012a}]{newman2012shocked}
Newman S.~F.,  et~al., 2012a, The Astrophysical Journal, 752, 111

\bibitem[\protect\citeauthoryear{Newman et~al.,}{Newman
  et~al.}{2012b}]{newman2012sins}
Newman S.~F.,  et~al., 2012b, The Astrophysical Journal, 761, 43

\bibitem[\protect\citeauthoryear{{Newville} et~al.,}{{Newville}
  et~al.}{2019}]{newville2019lmfit0.9.14}
{Newville} M.,  et~al., 2019, {lmfit/lmfit-py 0.9.14},
  \mn@doi{10.5281/zenodo.3381550}

\bibitem[\protect\citeauthoryear{{Oppenheimer} \& {Dav{\'e}}}{{Oppenheimer} \&
  {Dav{\'e}}}{2006}]{oppenheimer2006cosmologicalsims}
{Oppenheimer} B.~D.,  {Dav{\'e}} R.,  2006, \mn@doi [\mnras]
  {10.1111/j.1365-2966.2006.10989.x}, \href
  {https://ui.adsabs.harvard.edu/abs/2006MNRAS.373.1265O} {373, 1265}

\bibitem[\protect\citeauthoryear{{{\"O}stlin}, {Hayes}, {Kunth}, {Mas-Hesse},
  {Leitherer}, {Petrosian}  \& {Atek}}{{{\"O}stlin}
  et~al.}{2009}]{ostlin2009lymanalphamorphology}
{{\"O}stlin} G.,  {Hayes} M.,  {Kunth} D.,  {Mas-Hesse} J.~M.,  {Leitherer} C.,
   {Petrosian} A.,   {Atek} H.,  2009, \mn@doi [\aj]
  {10.1088/0004-6256/138/3/923}, \href
  {https://ui.adsabs.harvard.edu/abs/2009AJ....138..923O} {138, 923}

\bibitem[\protect\citeauthoryear{Ostriker, McKee  \& Leroy}{Ostriker
  et~al.}{2010}]{ostriker2010regulation}
Ostriker E.~C.,  McKee C.~F.,   Leroy A.~K.,  2010, The Astrophysical Journal,
  721, 975

\bibitem[\protect\citeauthoryear{{Ot{\'\i}-Floranes}, {Mas-Hesse},
  {Jim{\'e}nez-Bail{\'o}n}, {Schaerer}, {Hayes}, {{\"O}stlin}, {Atek}  \&
  {Kunth}}{{Ot{\'\i}-Floranes}
  et~al.}{2014}]{otifloranes2014physicalpropertiesIRAS08}
{Ot{\'\i}-Floranes} H.,  {Mas-Hesse} J.~M.,  {Jim{\'e}nez-Bail{\'o}n} E.,
  {Schaerer} D.,  {Hayes} M.,  {{\"O}stlin} G.,  {Atek} H.,   {Kunth} D.,
  2014, \mn@doi [\aap] {10.1051/0004-6361/201323069}, \href
  {https://ui.adsabs.harvard.edu/abs/2014A&A...566A..38O} {566, A38}

\bibitem[\protect\citeauthoryear{{Roberts-Borsani}}{{Roberts-Borsani}}{2020}]{robertsborsani2020observational}
{Roberts-Borsani} G.~W.,  2020, \mn@doi [\mnras] {10.1093/mnras/staa1006},
  \href {https://ui.adsabs.harvard.edu/abs/2020MNRAS.494.4266R} {494, 4266}

\bibitem[\protect\citeauthoryear{{Roberts-Borsani}, {Saintonge}, {Masters}  \&
  {Stark}}{{Roberts-Borsani} et~al.}{2020}]{robertsborsani2020outflowsMANGA}
{Roberts-Borsani} G.~W.,  {Saintonge} A.,  {Masters} K.~L.,   {Stark} D.~V.,
  2020, \mn@doi [\mnras] {10.1093/mnras/staa464}, \href
  {https://ui.adsabs.harvard.edu/abs/2020MNRAS.tmp..439R} {}

\bibitem[\protect\citeauthoryear{{Rodr{\'\i}guez del Pino}, {Arribas},
  {Piqueras L{\'o}pez}, {Villar-Mart{\'\i}n}  \& {Colina}}{{Rodr{\'\i}guez del
  Pino} et~al.}{2019}]{RodriguezdelPino2019ionisedoutflowsmanga}
{Rodr{\'\i}guez del Pino} B.,  {Arribas} S.,  {Piqueras L{\'o}pez} J.,
  {Villar-Mart{\'\i}n} M.,   {Colina} L.,  2019, \mn@doi [\mnras]
  {10.1093/mnras/stz816}, \href
  {https://ui.adsabs.harvard.edu/abs/2019MNRAS.486..344R} {486, 344}

\bibitem[\protect\citeauthoryear{{Rubin}, {Prochaska}, {Koo}, {Phillips},
  {Martin}  \& {Winstrom}}{{Rubin} et~al.}{2014}]{Rubin2014evidence}
{Rubin} K. H.~R.,  {Prochaska} J.~X.,  {Koo} D.~C.,  {Phillips} A.~C.,
  {Martin} C.~L.,   {Winstrom} L.~O.,  2014, \mn@doi [\apj]
  {10.1088/0004-637X/794/2/156}, \href
  {https://ui.adsabs.harvard.edu/abs/2014ApJ...794..156R} {794, 156}

\bibitem[\protect\citeauthoryear{{Rupke}, {Veilleux}  \& {Sanders}}{{Rupke}
  et~al.}{2005}]{rupke2005outflowsdiscussion}
{Rupke} D.~S.,  {Veilleux} S.,   {Sanders} D.~B.,  2005, \mn@doi [\apjs]
  {10.1086/432889}, \href
  {https://ui.adsabs.harvard.edu/abs/2005ApJS..160..115R} {160, 115}

\bibitem[\protect\citeauthoryear{{Saintonge} et~al.,}{{Saintonge}
  et~al.}{2017}]{saintonge2017xColdGass}
{Saintonge} A.,  et~al., 2017, \mn@doi [\apjs] {10.3847/1538-4365/aa97e0},
  \href {https://ui.adsabs.harvard.edu/abs/2017ApJS..233...22S} {233, 22}

\bibitem[\protect\citeauthoryear{{Shetty} \& {Ostriker}}{{Shetty} \&
  {Ostriker}}{2012}]{Shetty2012maximally}
{Shetty} R.,  {Ostriker} E.~C.,  2012, \mn@doi [\apj]
  {10.1088/0004-637X/754/1/2}, \href
  {https://ui.adsabs.harvard.edu/abs/2012ApJ...754....2S} {754, 2}

\bibitem[\protect\citeauthoryear{{Shopbell} \& {Bland-Hawthorn}}{{Shopbell} \&
  {Bland-Hawthorn}}{1998}]{shopbell1998asymmetricwindM82}
{Shopbell} P.~L.,  {Bland-Hawthorn} J.,  1998, \mn@doi [\apj] {10.1086/305108},
  \href {https://ui.adsabs.harvard.edu/abs/1998ApJ...493..129S} {493, 129}

\bibitem[\protect\citeauthoryear{{Springel} \& {Hernquist}}{{Springel} \&
  {Hernquist}}{2003}]{springel2003cosmologicalsim}
{Springel} V.,  {Hernquist} L.,  2003, \mn@doi [\mnras]
  {10.1046/j.1365-8711.2003.06206.x}, \href
  {https://ui.adsabs.harvard.edu/abs/2003MNRAS.339..289S} {339, 289}

\bibitem[\protect\citeauthoryear{{Stanway} \& {Eldridge}}{{Stanway} \&
  {Eldridge}}{2018}]{stanway2018reevaluatingbpass}
{Stanway} E.~R.,  {Eldridge} J.~J.,  2018, \mn@doi [\mnras]
  {10.1093/mnras/sty1353}, \href
  {https://ui.adsabs.harvard.edu/abs/2018MNRAS.479...75S} {479, 75}

\bibitem[\protect\citeauthoryear{{Swinbank} et~al.,}{{Swinbank}
  et~al.}{2019}]{swinbank2019energetics}
{Swinbank} A.~M.,  et~al., 2019, \mn@doi [\mnras] {10.1093/mnras/stz1275},
  \href {https://ui.adsabs.harvard.edu/abs/2019MNRAS.487..381S} {487, 381}

\bibitem[\protect\citeauthoryear{{Tacconi} et~al.,}{{Tacconi}
  et~al.}{2018}]{tacconi2018PHIBSSunified}
{Tacconi} L.~J.,  et~al., 2018, \mn@doi [\apj] {10.3847/1538-4357/aaa4b4},
  \href {https://ui.adsabs.harvard.edu/abs/2018ApJ...853..179T} {853, 179}

\bibitem[\protect\citeauthoryear{{Tumlinson}, {Peeples}  \& {Werk}}{{Tumlinson}
  et~al.}{2017}]{tumlinson2017CGMreview}
{Tumlinson} J.,  {Peeples} M.~S.,   {Werk} J.~K.,  2017, \mn@doi [\araa]
  {10.1146/annurev-astro-091916-055240}, \href
  {https://ui.adsabs.harvard.edu/abs/2017ARA&A..55..389T} {55, 389}

\bibitem[\protect\citeauthoryear{{{\"U}bler} et~al.,}{{{\"U}bler}
  et~al.}{2019}]{ubler2019evolutionandorigin}
{{\"U}bler} H.,  et~al., 2019, \mn@doi [\apj] {10.3847/1538-4357/ab27cc}, \href
  {https://ui.adsabs.harvard.edu/abs/2019ApJ...880...48U} {880, 48}

\bibitem[\protect\citeauthoryear{{Veilleux} \& {Rupke}}{{Veilleux} \&
  {Rupke}}{2002}]{veilleux2002identificationNGC1482}
{Veilleux} S.,  {Rupke} D.~S.,  2002, \mn@doi [\apjl] {10.1086/339226}, \href
  {https://ui.adsabs.harvard.edu/abs/2002ApJ...565L..63V} {565, L63}

\bibitem[\protect\citeauthoryear{{Veilleux}, {Cecil}  \&
  {Bland-Hawthorn}}{{Veilleux} et~al.}{2005}]{veilleux2005galacticwinds}
{Veilleux} S.,  {Cecil} G.,   {Bland-Hawthorn} J.,  2005, \mn@doi [\araa]
  {10.1146/annurev.astro.43.072103.150610}, \href
  {https://ui.adsabs.harvard.edu/abs/2005ARA&A..43..769V} {43, 769}

\bibitem[\protect\citeauthoryear{{Veilleux}, {Maiolino}, {Bolatto}  \&
  {Aalto}}{{Veilleux} et~al.}{2020}]{veilleux2020cooloutflows}
{Veilleux} S.,  {Maiolino} R.,  {Bolatto} A.~D.,   {Aalto} S.,  2020, \mn@doi
  [\aapr] {10.1007/s00159-019-0121-9}, \href
  {https://ui.adsabs.harvard.edu/abs/2020A&ARv..28....2V} {28, 2}

\bibitem[\protect\citeauthoryear{Weiner et~al.,}{Weiner
  et~al.}{2009}]{weiner2009ubiquitous}
Weiner B.~J.,  et~al., 2009, The Astrophysical Journal, 692, 187

\bibitem[\protect\citeauthoryear{{Wilson}, {Elmegreen}, {Bemis}  \&
  {Brunetti}}{{Wilson} et~al.}{2019}]{wilson2019kennicuttschmidtlaw}
{Wilson} C.~D.,  {Elmegreen} B.~G.,  {Bemis} A.,   {Brunetti} N.,  2019,
  \mn@doi [\apj] {10.3847/1538-4357/ab31f3}, \href
  {https://ui.adsabs.harvard.edu/abs/2019ApJ...882....5W} {882, 5}

\bibitem[\protect\citeauthoryear{{Wisnioski} et~al.,}{{Wisnioski}
  et~al.}{2015}]{wisnioski2015KMOS3Dsurvey}
{Wisnioski} E.,  et~al., 2015, \mn@doi [\apj] {10.1088/0004-637X/799/2/209},
  \href {https://ui.adsabs.harvard.edu/abs/2015ApJ...799..209W} {799, 209}

\makeatother
\end{thebibliography}






\bsp	
\label{lastpage}
\end{document}